\documentclass[%
reprint,
superscriptaddress,
bibnotes,
amsmath,amssymb,
aps,
pra,
showkeys,
]{revtex4-2}

\usepackage{graphicx}
\usepackage[caption=false]{subfig}
\usepackage{qcircuit}
\usepackage{dcolumn}
\usepackage{bm}
\usepackage{tikz}
\usepackage{mathtools}
\usetikzlibrary{plotmarks}
\usetikzlibrary{decorations.markings}
\usepackage{siunitx}
\usepackage{xy}
\usepackage{dsfont}
\usepackage{gensymb}
\usepackage{pgfplots}
\usepackage{pgf}
\pgfplotsset{compat=1.15}
\usepackage{placeins}
\usepackage{flafter}
\usepackage{color}
\usepackage[normalem]{ulem} 
\usepackage{soul} 
\usepackage{cancel} 
\usepackage{orcidlink}
\usepackage{hyperref}
\hypersetup{
	colorlinks=true,
	linkcolor=blue,
	filecolor=blue,      
	urlcolor=magenta,
	citecolor=blue,
	pdftex,
	pdfauthor={Ahmed M. Farouk, I. I. Beterov, Ghadeer Suliman, Junxi Chen, and I.I. Ryabtsev},
	pdftitle={},
   pdfcreator={\LaTeX~by Ahmed M. Farouk},
}
\usetikzlibrary{math} 
\usepackage{epstopdf}
\makeatletter
\newcommand\tinyv{\@setfontsize\tinyv{5pt}{5}}
\makeatother

\tikzmath{
\iconSSW=2.3;
\iconSW=1.85;
\iconKW=1.75;
\iconEKW=1.5*\iconKW;
\iconFKW=2*\iconKW;
\iconString=2.5*\iconKW;
}
\newcommand*\iconTriangleA{\raisebox{-0.45\baselineskip}{\includegraphics[width=\iconSW\baselineskip]{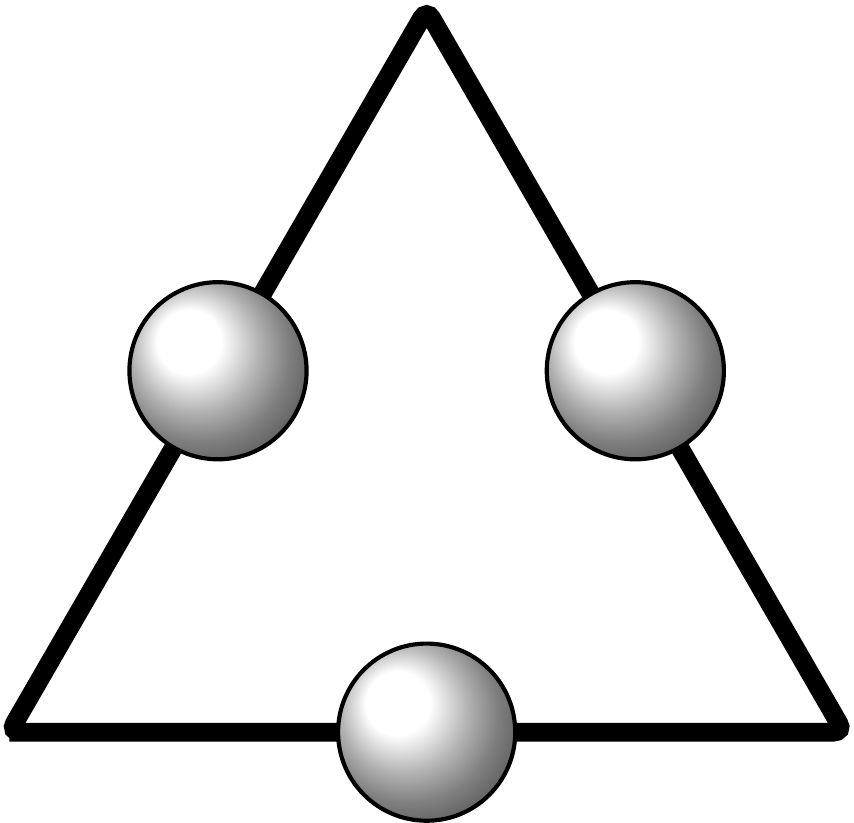}}~}
\newcommand*\iconTriangleB{\raisebox{-0.45\baselineskip}{\includegraphics[width=\iconSW\baselineskip]{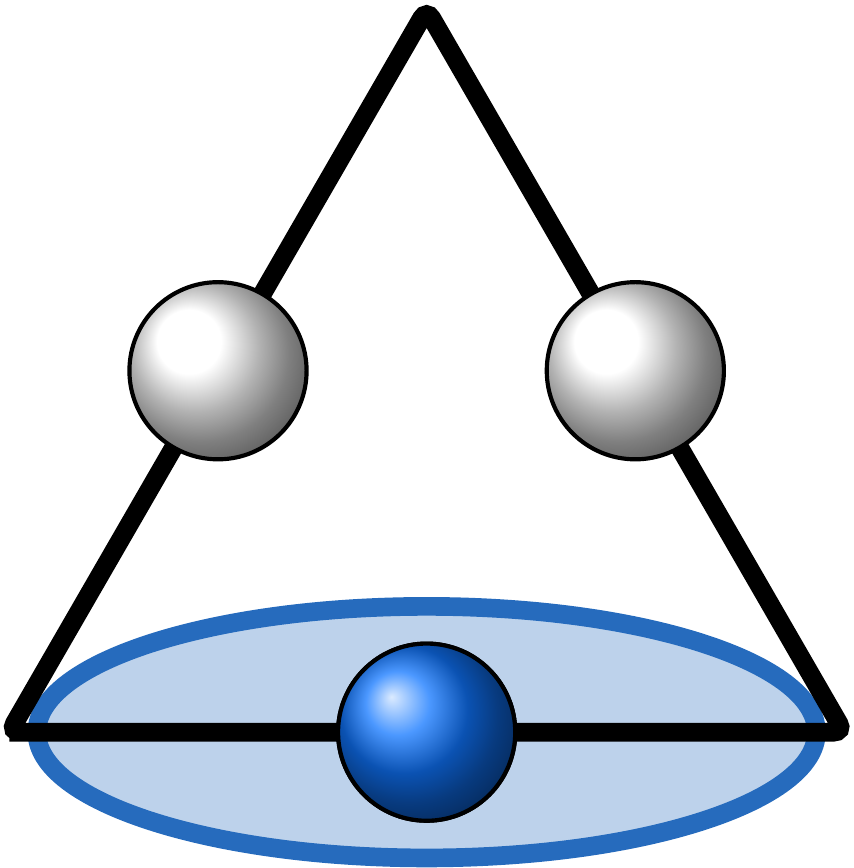}}~}
\newcommand*\iconTriangleC{\raisebox{-0.45\baselineskip}{\includegraphics[width=\iconSW\baselineskip]{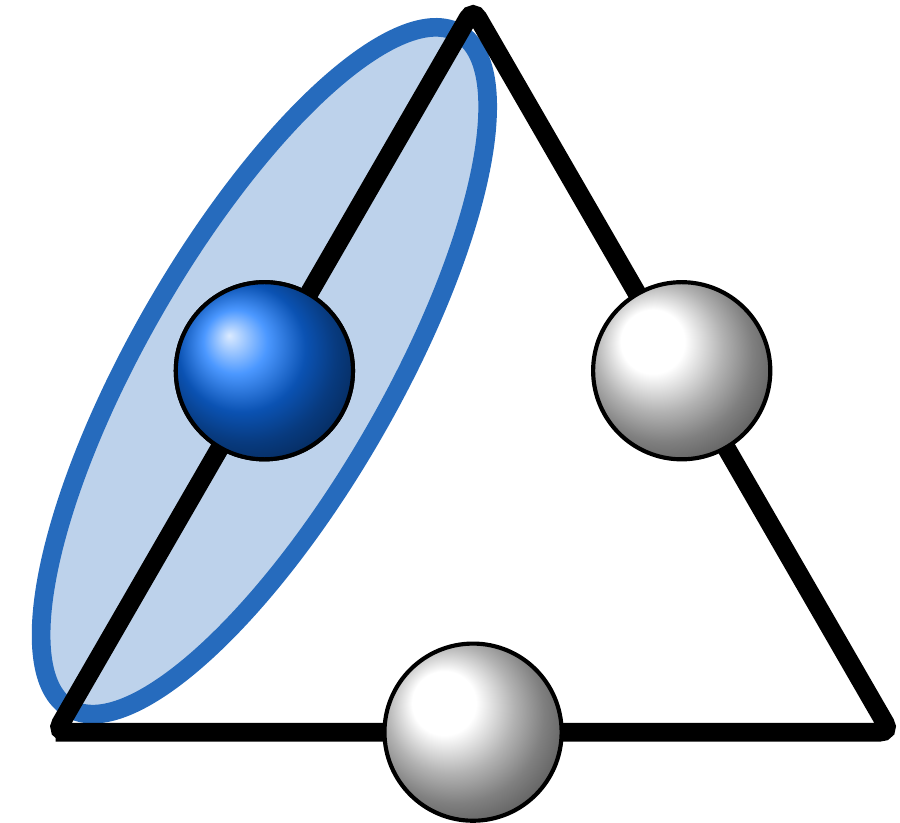}}~}
\newcommand*\iconTriangleD{\raisebox{-0.45\baselineskip}{\includegraphics[width=\iconSW\baselineskip]{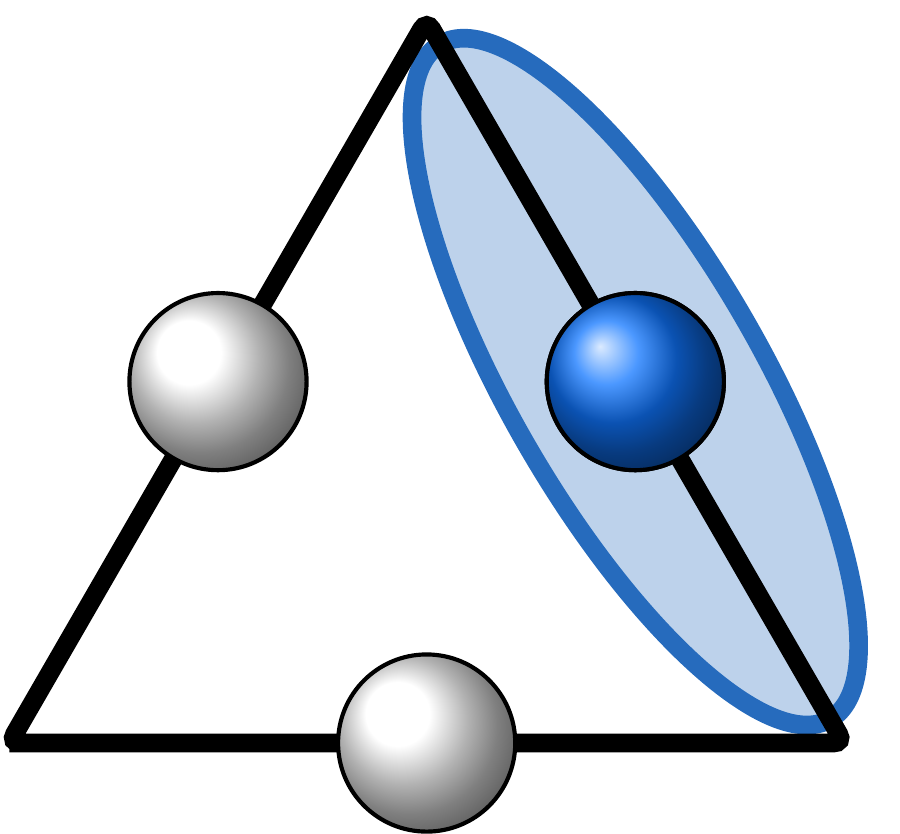}}~}
\newcommand*\iconTriangleStringZ{\raisebox{-0.45\baselineskip}{\includegraphics[width=\iconSW\baselineskip]{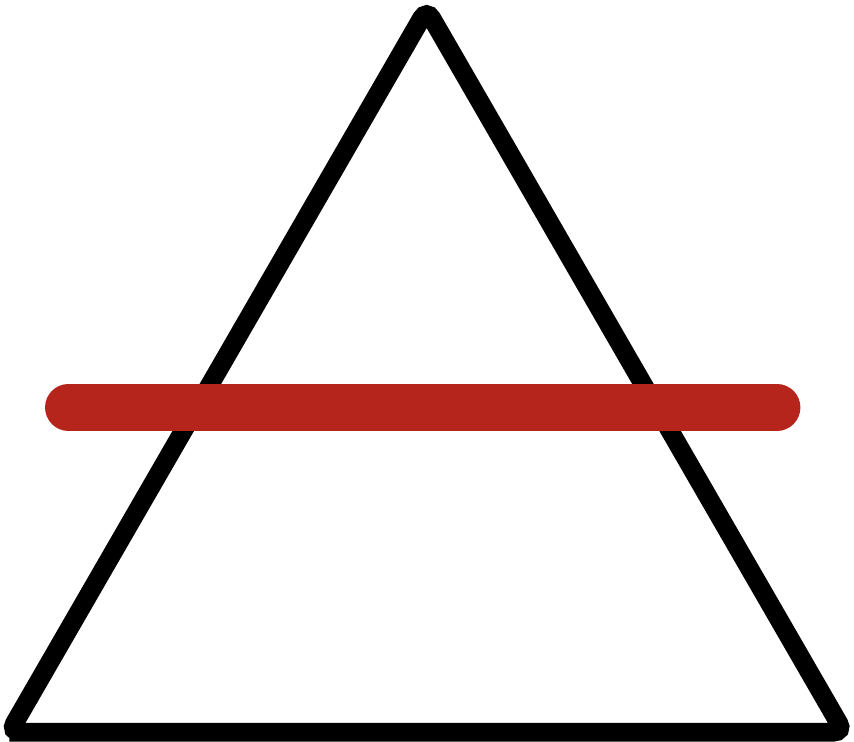}}~}
\newcommand*\iconTriangleStringX{\raisebox{-0.45\baselineskip}{\includegraphics[width=\iconSW\baselineskip]{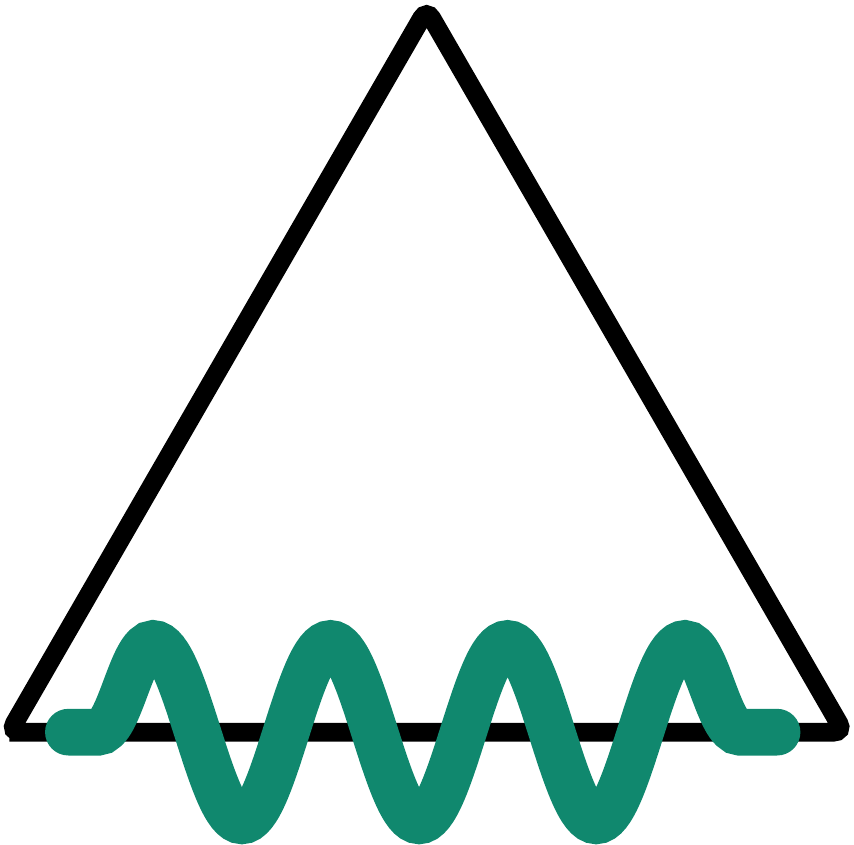}}~}
\newcommand*\iconUnitCell{\raisebox{-0.45\baselineskip}{\includegraphics[width=\iconSW\baselineskip]{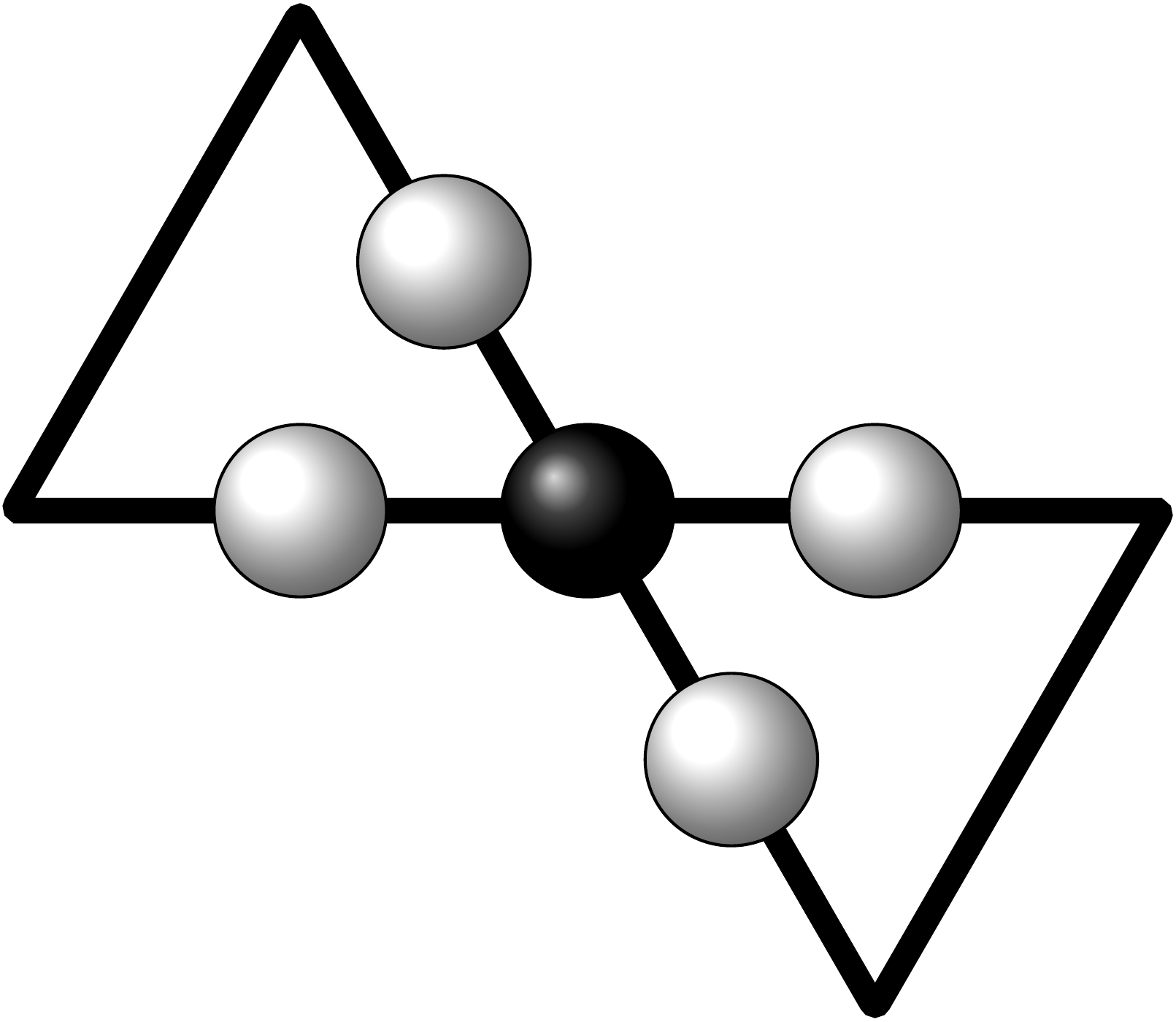}}~}
\newcommand*\iconMonomer{\raisebox{-0.45\baselineskip}{\includegraphics[width=\iconSW\baselineskip]{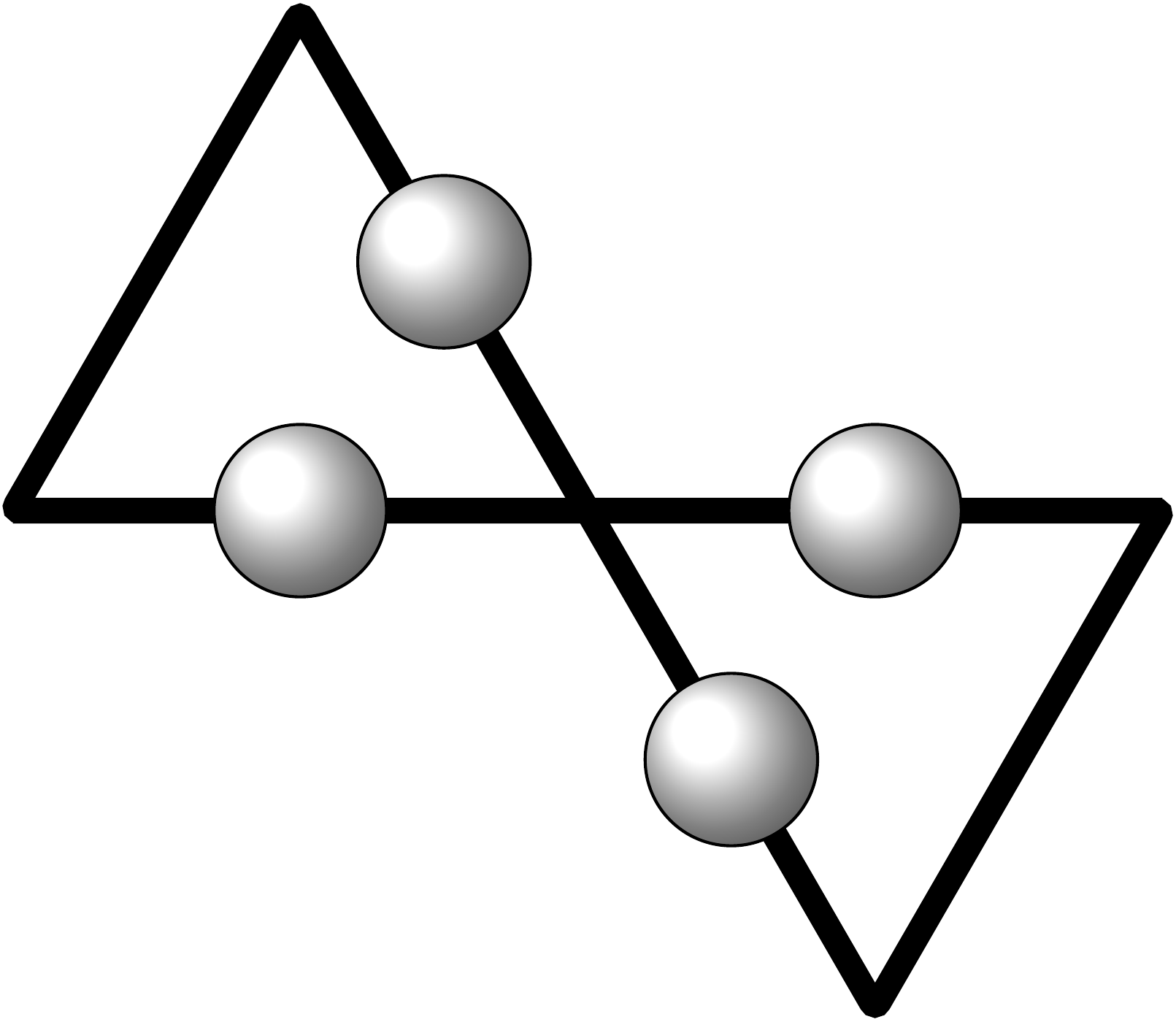}}~}
\newcommand*\iconDimer{\raisebox{-0.45\baselineskip}{\includegraphics[width=\iconSW\baselineskip]{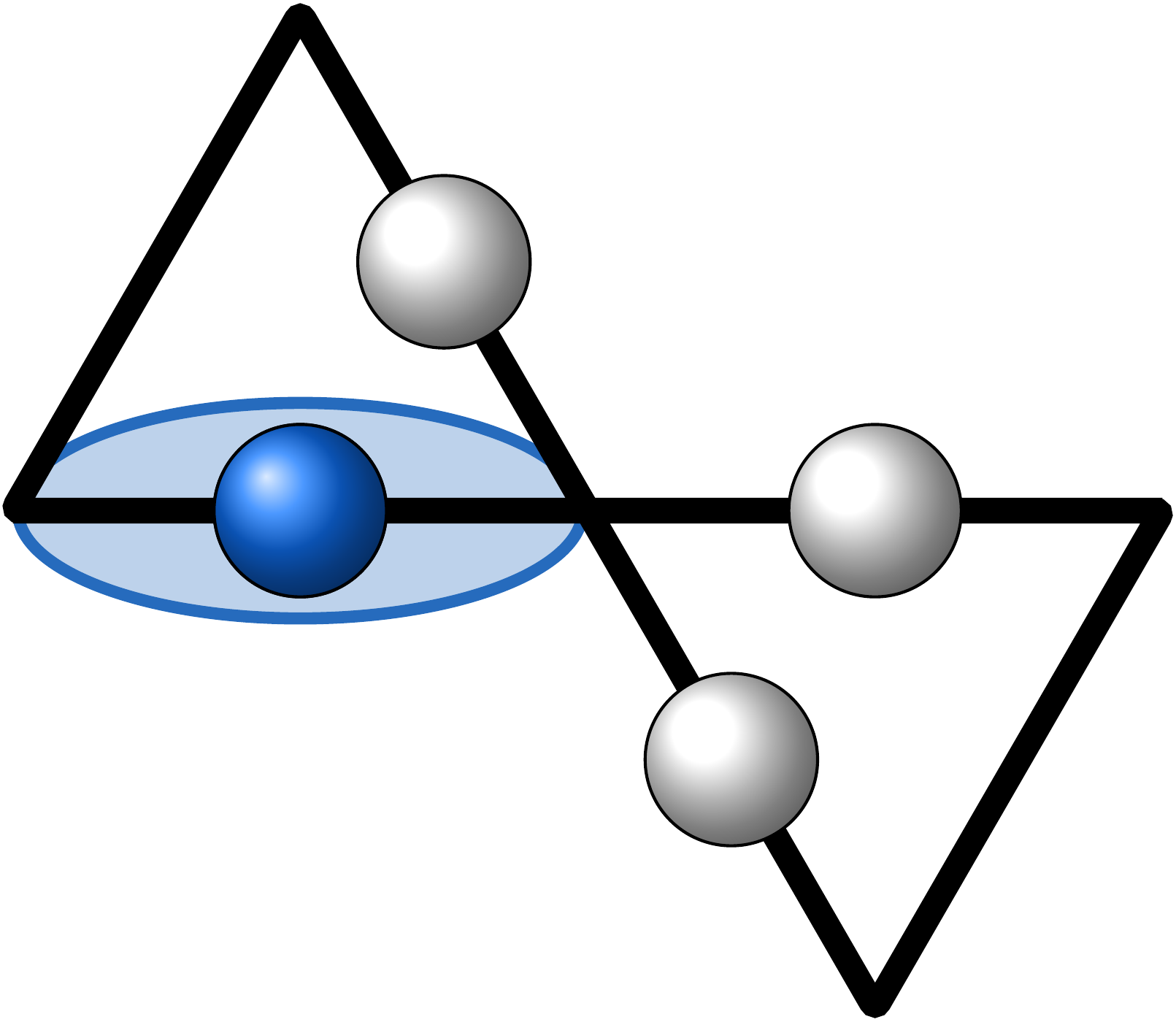}}~}
\newcommand*\iconDimerDouble{\raisebox{-0.45\baselineskip}{\includegraphics[width=\iconSW\baselineskip]{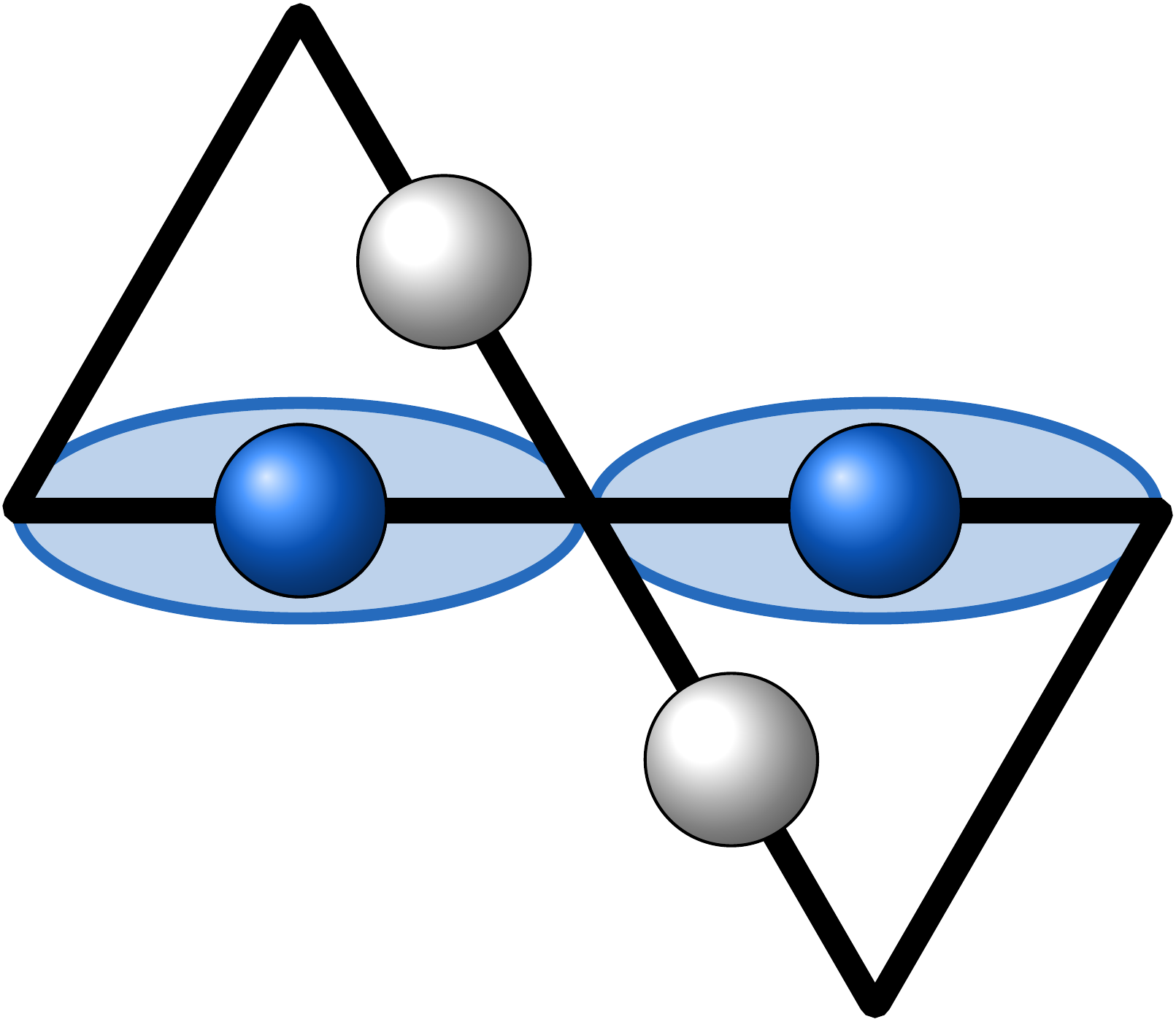}}~}
\newcommand*\iconDimerDoubleB{\raisebox{-0.45\baselineskip}{\includegraphics[width=\iconSW\baselineskip]{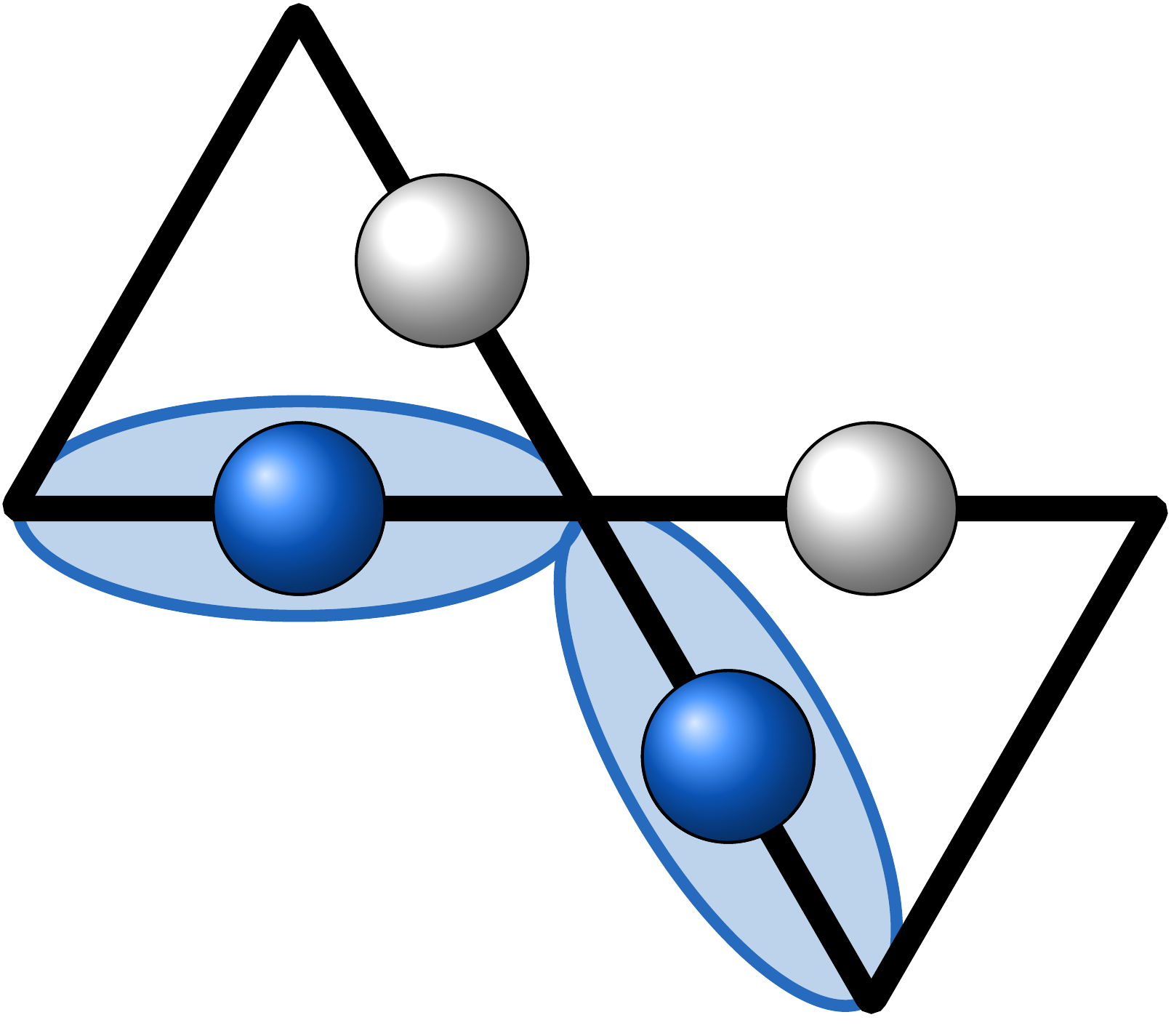}}~}
\newcommand*\iconDimerMonomerNN{\raisebox{-0.2\baselineskip}{\includegraphics[width=\iconSSW\baselineskip]{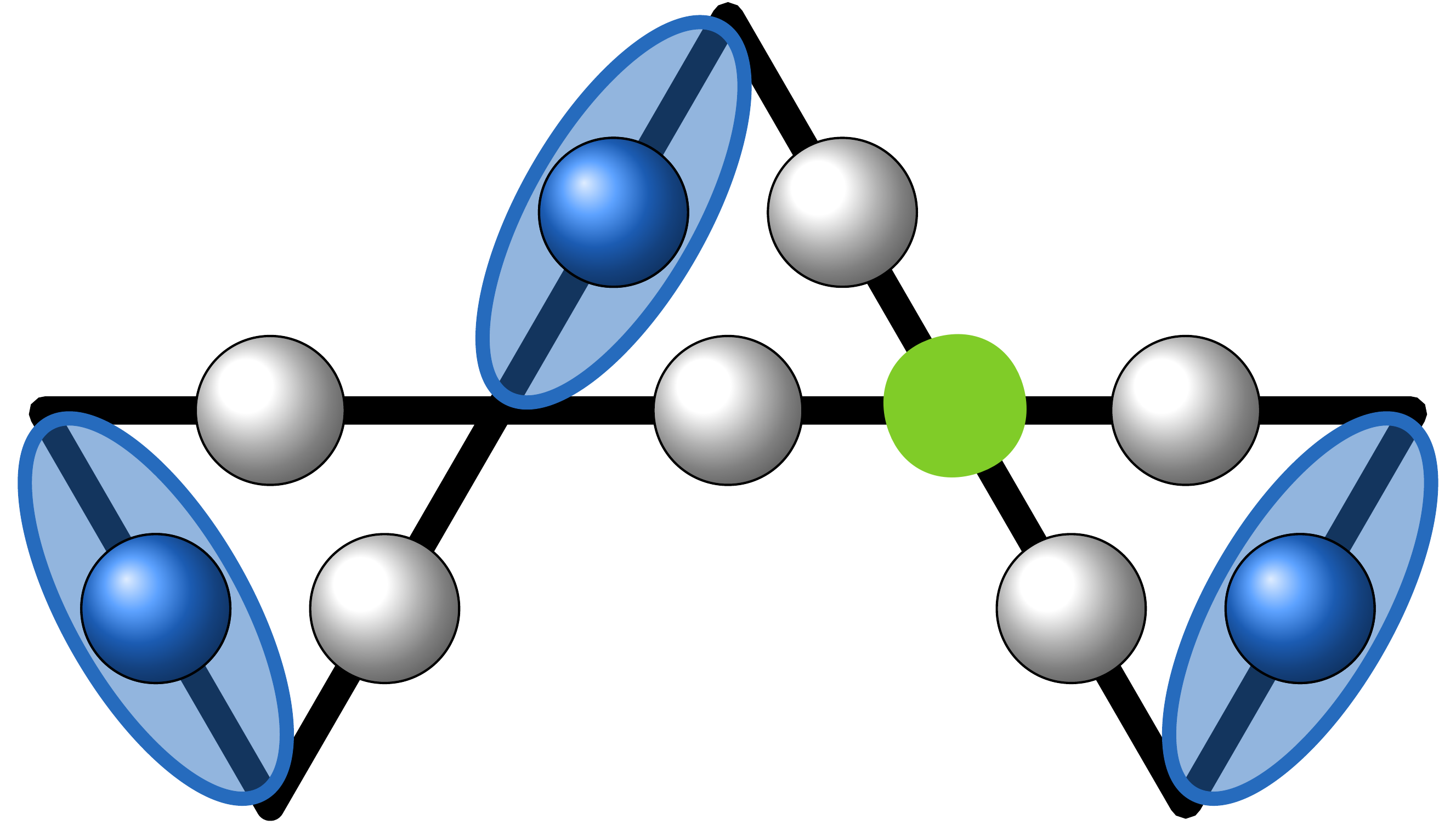}}~}
\newcommand*\iconDimerNN{\raisebox{-0.2\baselineskip}{\includegraphics[width=\iconSSW\baselineskip]{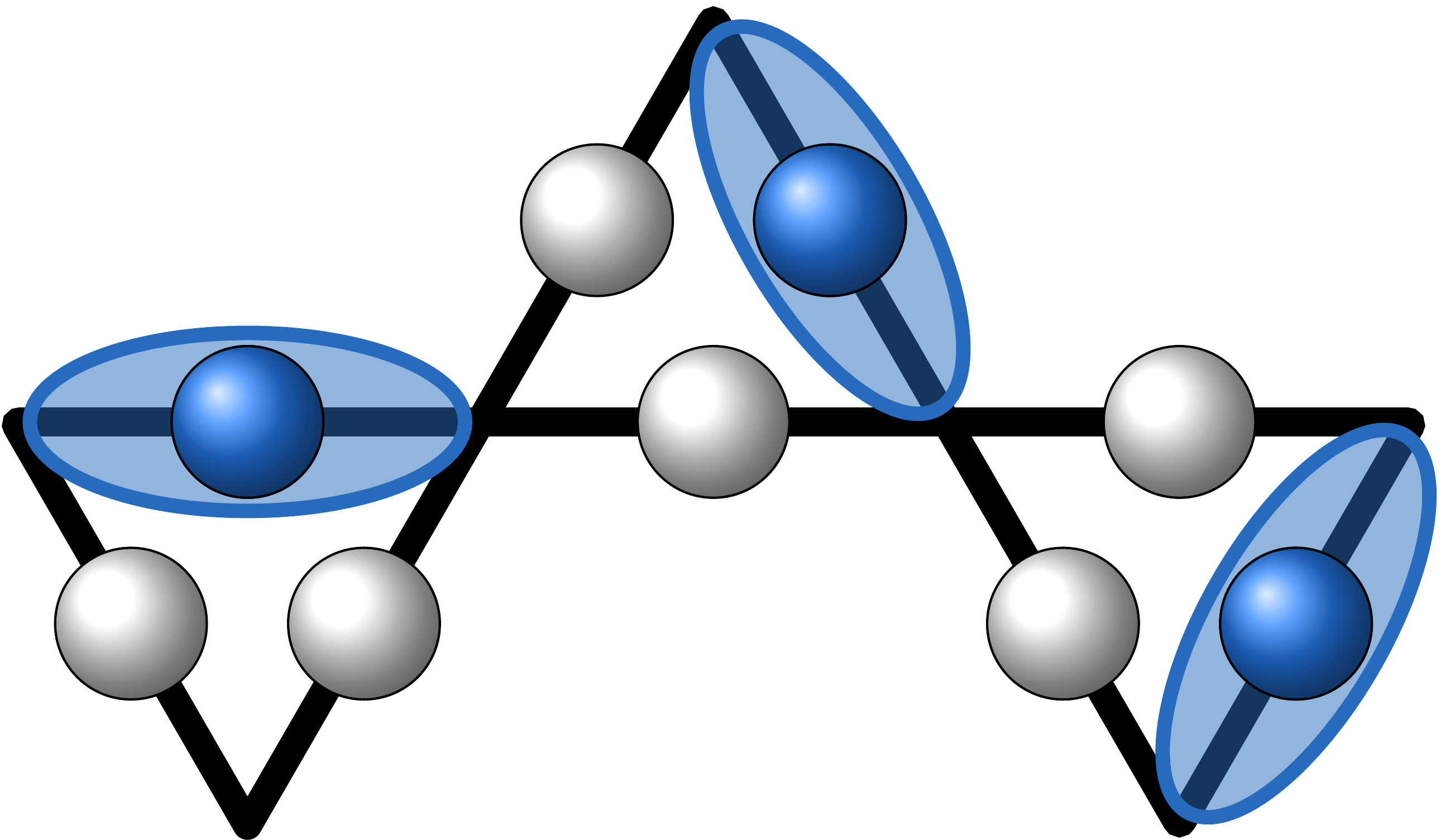}}~}
\newcommand*\iconKagome{\raisebox{-0.65\baselineskip}{\includegraphics[width=\iconKW\baselineskip]{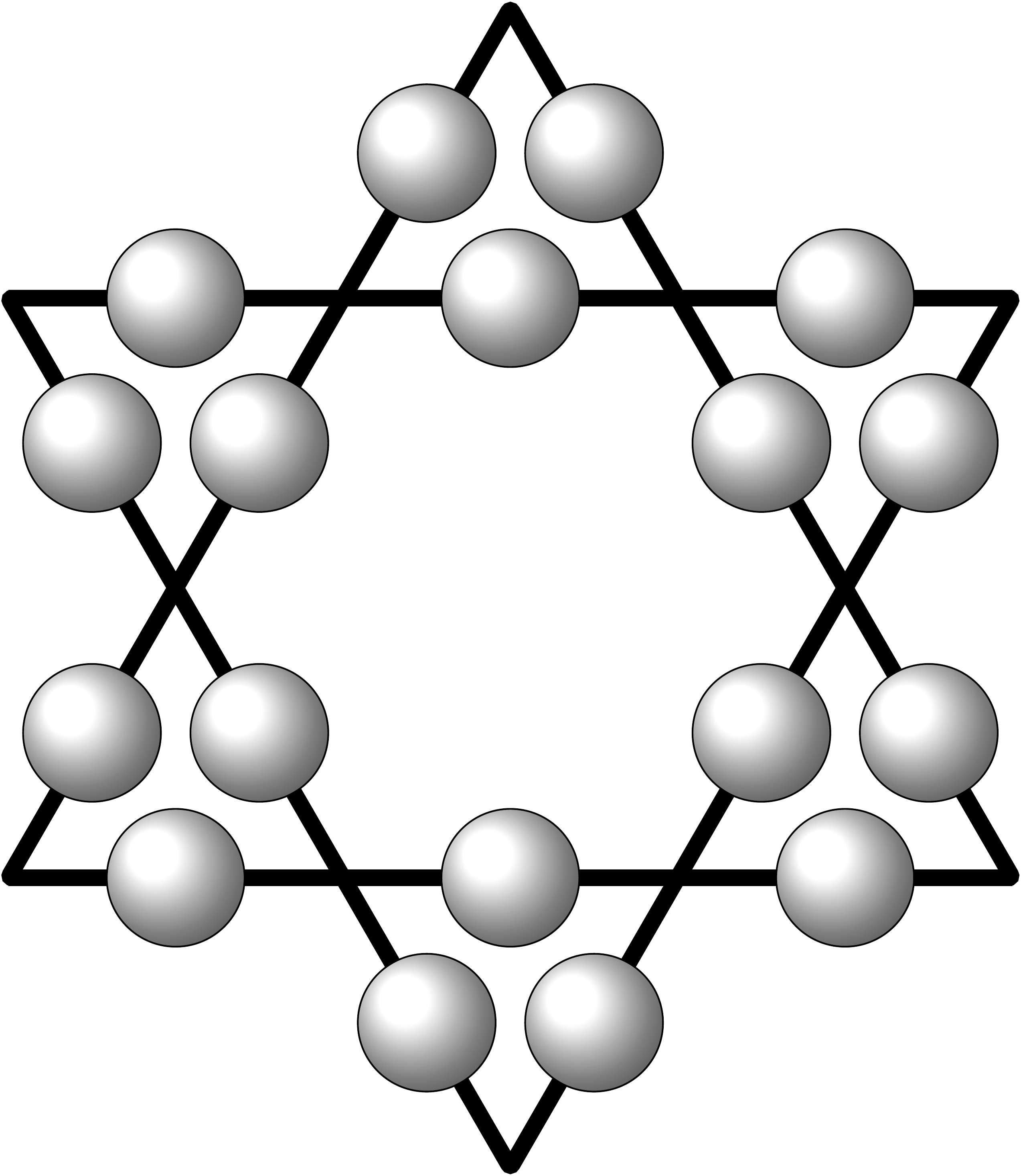}}~}

\newcommand*\iconEnKagomeMMMDimerA{\raisebox{-0.85\baselineskip}{\includegraphics[width=\iconEKW\baselineskip]{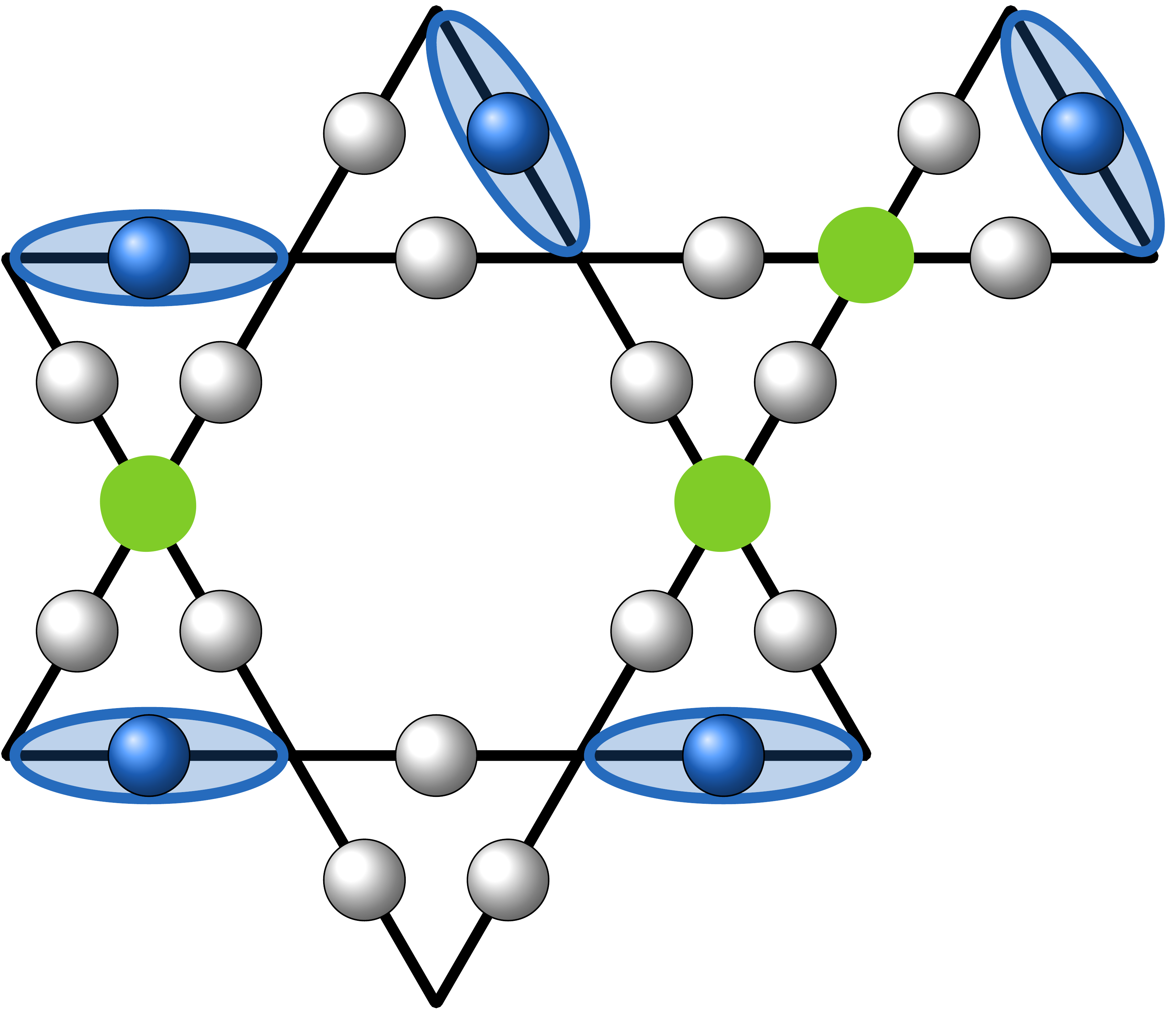}}~}
\newcommand*\iconEnKagomeMMDimerA{\raisebox{-0.85\baselineskip}{\includegraphics[width=\iconEKW\baselineskip]{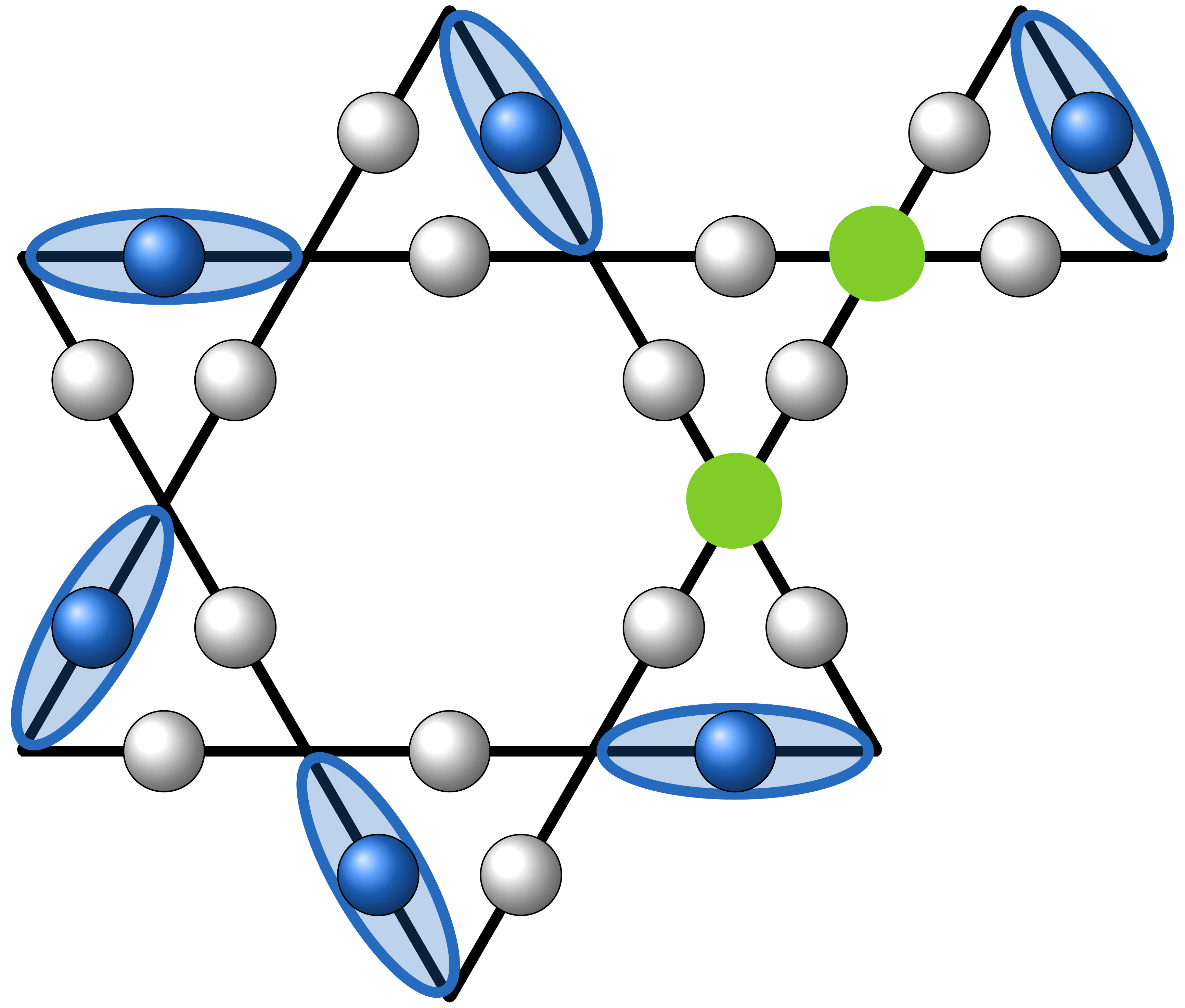}}~}
\newcommand*\iconEnKagomeMDimerA{\raisebox{-0.85\baselineskip}{\includegraphics[width=\iconEKW\baselineskip]{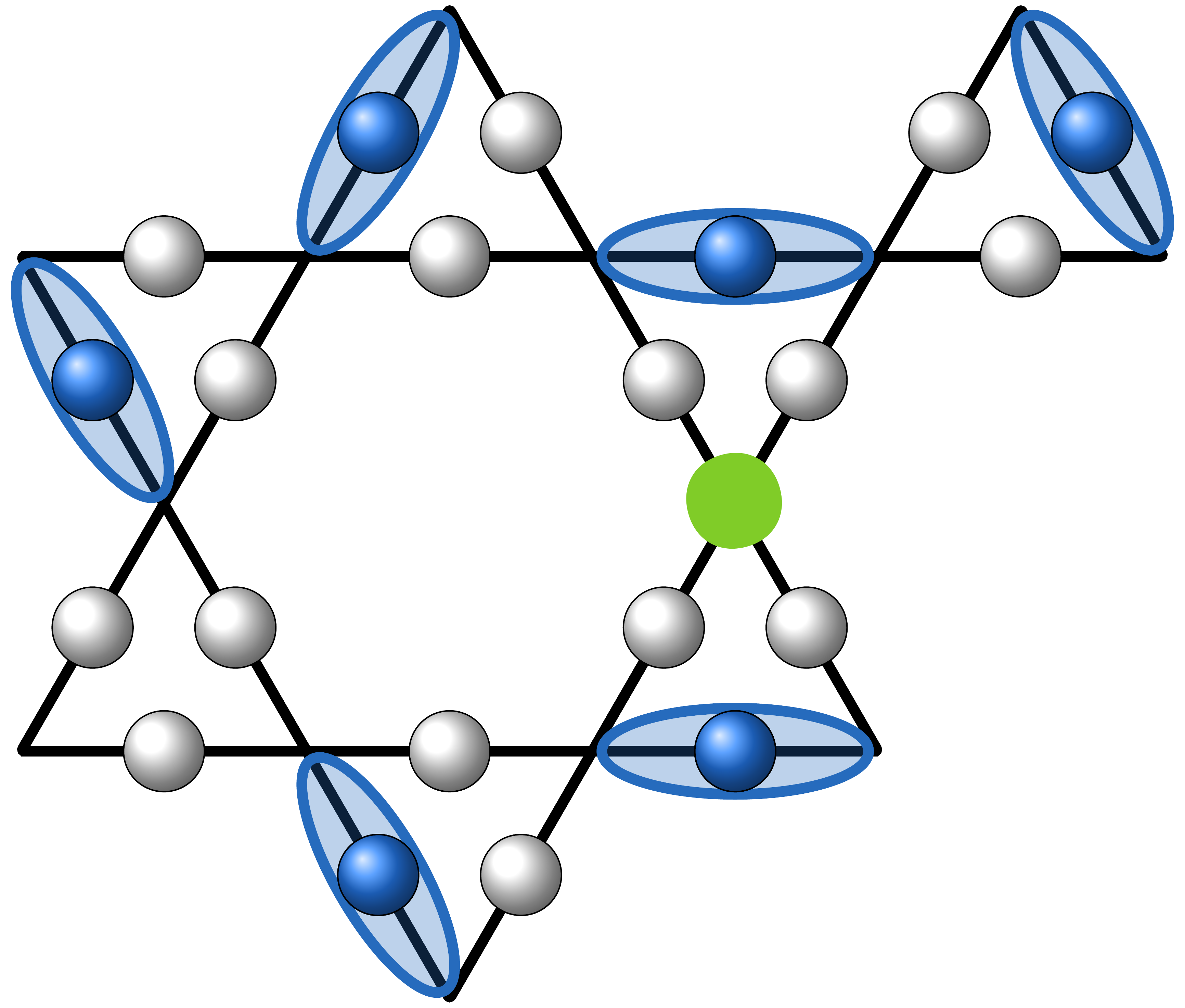}}~}

\newcommand*\iconFKagomeA{\raisebox{-0.95\baselineskip}{\includegraphics[width=\iconFKW\baselineskip]{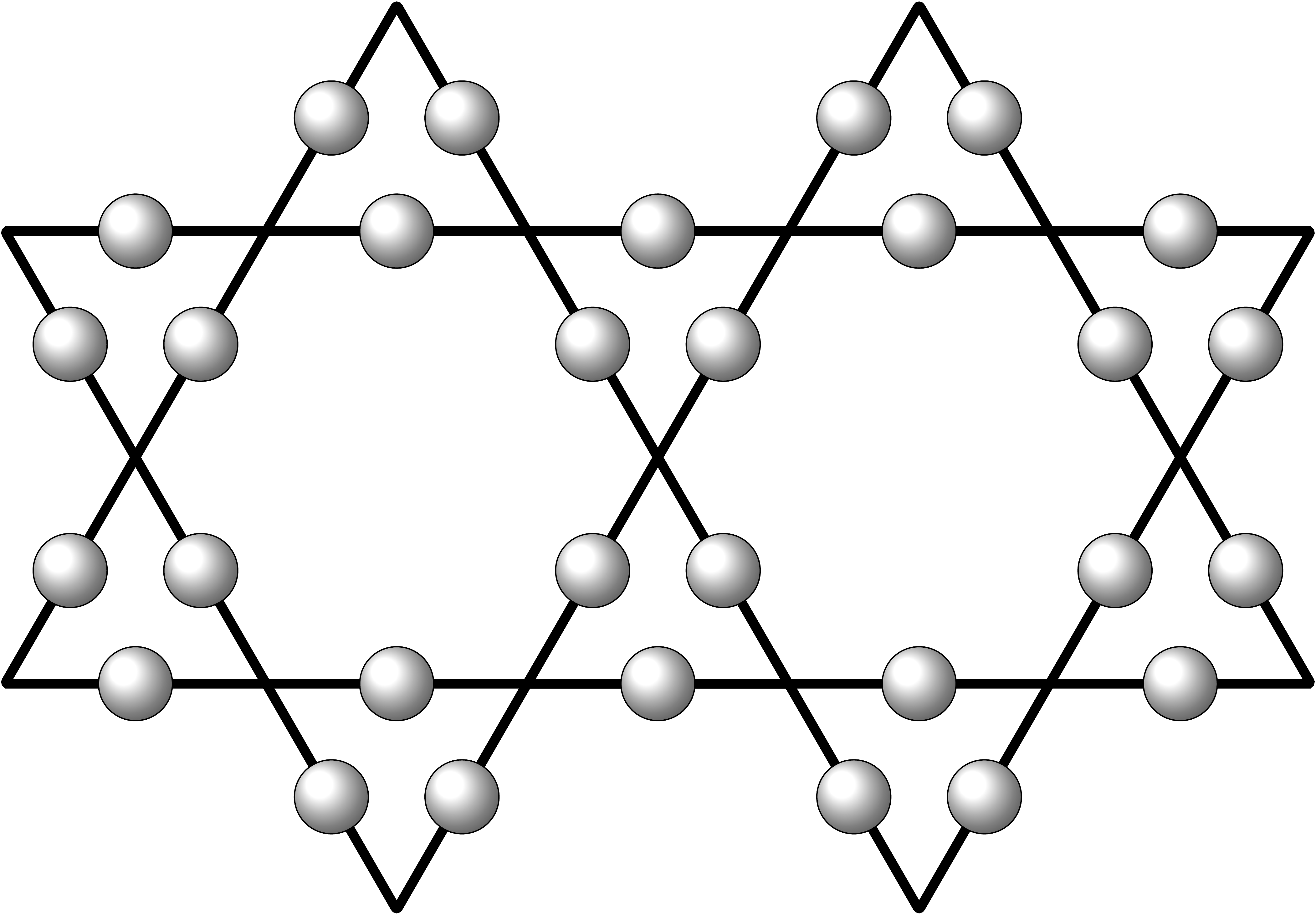}}~}
\begin{document}


\title{Quantum Spin Liquid State of a Dual-Species Atomic Array on Kagome Lattice} 
\author{Ahmed M. Farouk~\orcidlink{0000-0002-6230-1234}}
\email{afarouk@isp.nsc.ru}
\affiliation{Laboratory of Nonlinear Resonant Processes and Laser Diagnostics, Rzhanov Institute of Semiconductor Physics SB RAS, 630090 Novosibirsk, Russia}
\author{Ilya I. Beterov~\orcidlink{0000-0002-6596-6741}}
\email{beterov@isp.nsc.ru}
\affiliation{Laboratory of Nonlinear Resonant Processes and Laser Diagnostics, Rzhanov Institute of Semiconductor Physics SB RAS, 630090 Novosibirsk, Russia}
\affiliation{Faculty of Physics, Novosibirsk State University, 630090 Novosibirsk, Russia}
\affiliation{Department of Laser Physics, Institute of Laser Physics SB RAS, 630090 Novosibirsk, Russia}
\affiliation{Faculty of Physical Engineering, Novosibirsk State Technical University, 630073 Novosibirsk, Russia}
\author{Ghadeer Suliman~\orcidlink{0009-0000-3769-4235}}
\affiliation{Laboratory of Nonlinear Resonant Processes and Laser Diagnostics, Rzhanov Institute of Semiconductor Physics SB RAS, 630090 Novosibirsk, Russia}
\affiliation{Faculty of Physics, Novosibirsk State University, 630090 Novosibirsk, Russia}
\author{Junxi Chen~\orcidlink{0009-0001-2107-2543}}
\affiliation{Laboratory of Nonlinear Resonant Processes and Laser Diagnostics, Rzhanov Institute of Semiconductor Physics SB RAS, 630090 Novosibirsk, Russia}
\affiliation{Faculty of Physics, Novosibirsk State University, 630090 Novosibirsk, Russia}

\author{Igor I. Ryabtsev~\orcidlink{0000-0002-5410-2155}}
\affiliation{Laboratory of Nonlinear Resonant Processes and Laser Diagnostics, Rzhanov Institute of Semiconductor Physics SB RAS, 630090 Novosibirsk, Russia}
\affiliation{Faculty of Physics, Novosibirsk State University, 630090 Novosibirsk, Russia}

\date{\today}
\begin{abstract}
Dual-species arrays of ultracold neutral atoms have recently attracted increased interest due to the ability to independently control different atomic species and tune the interatomic interactions. This capability provides additional flexibility essential for both quantum computing and quantum simulation. In this work we theoretically investigate a quantum spin liquid (QSL) state to be simulated on a programmable quantum simulator based on a dual-species atomic array, arranged on a Kagome lattice.  The Kagome lattice is formed by corner sharing triangles. This specific spatial arrangement  enhances the competing interactions between atoms and is often considered as a model for realizing QSL states. When the atoms are excited into Rydberg states, long-range interactions result in Rydberg blockade. The geometric frustration of the Kagome lattice, combined with the Rydberg blockade, drives the system into exotic phases with topological order and long-range entanglement. To drive an array into the QSL state, we use a  sweep-hold-sweep protocol, when the atoms are quasi-adiabatically excited into Rydberg state with individually controlled detuning from the resonance for each atomic species. The filling fraction, indicating emergence of a QSL state, is represented by a density of Rydberg excitations. We identified the conditions required for QSL state in a dual-species array with non-uniform interaction energies.  We calculated the correlation length and studied the mutual information as a function of the size of the subset of the system. The existence of a topological order was proved by estimating the Kitaev-Preskill topological quantum entanglement entropy, and further confirmed by evaluating diagonal and off-diagonal topological string operators, which respectively probe the classical dimer-covering structure and the quantum coherence between distinct dimer configurations.
\end{abstract}

\maketitle

\section{Introduction}

Notable experimental advancements in the precise control of quantum states in large-scale atomic arrays have led to the development of new research tools, known as programmable quantum simulators~\cite{weimer2010rydberg, bernien2017probing,AltmanPRXQuantum2021, manovitz2025quantum}. Nowadays, ultracold atoms can be trapped in defect-free arrays of optical dipole traps with arbitrary spatial configurations~\cite{barredo2016atom}. Laser excitation of these atoms into Rydberg states generates entanglement due to long-range Rydberg interactions~\cite{SaffmanWalkerMolmer2013, browaeys2020many}. As a result, the quantum states of these atomic arrays mimic various quantum phases of matter~\cite{ebadi2021quantum}. These simulators enable finding solutions of complex problems in condensed-matter physics~\cite{gonzalez2025observation}, including magnetism~\cite{leclerc2026one}, quantum phase transitions, frustration, and numerous other phenomena that are challenging to simulate on classical computers~\cite{sachdev2023quantum}.

Long-range interactions in atomic arrays give rise to the Rydberg blockade~\cite{JakschCirac2000}, which prevents the simultaneous excitation of two nearby atoms into Rydberg states by resonant laser radiation. This effect is widely used to create entanglement in atomic arrays and to implement multi-qubit gates for quantum computing~\cite{IsenhowerPRL2010, LevinePRL2019, CongLevine2021, McDonnellPRL2022}. Beyond these applications, it provides opportunities to explore quantum phases of matter in diverse spatial configurations. The critical dynamics of topologically ordered states in one-dimensional~\cite{samajdar2018numerical, keesling2019quantum} and two-dimensional~\cite{Samajdar2020Complex, scholl2021quantum, fang2025probing, HomeierPRA2025} atomic arrays have been studied both theoretically and experimentally. Commensurate quantum phases of matter are particularly useful for implementing quantum algorithms to solve graph theory problems, including the maximum independent set~\cite{ByunPRX2022, kim2022rydberg}, maximum-weight independent set~\cite{farouk2024generation, lanthaler2024quantum, BombieriPRX2025, kombe2025quantum}, graph coloring~\cite{angkhanawin2025graph}, and max-cut problems~\cite{graham2022multi}. These problems have broad relevance across scientific fields and industry sectors~\cite{bauer2024solving, barreto2024, grotti2025practical}.

\begin{figure*}[t]
	\centering
	\includegraphics[width=\textwidth]{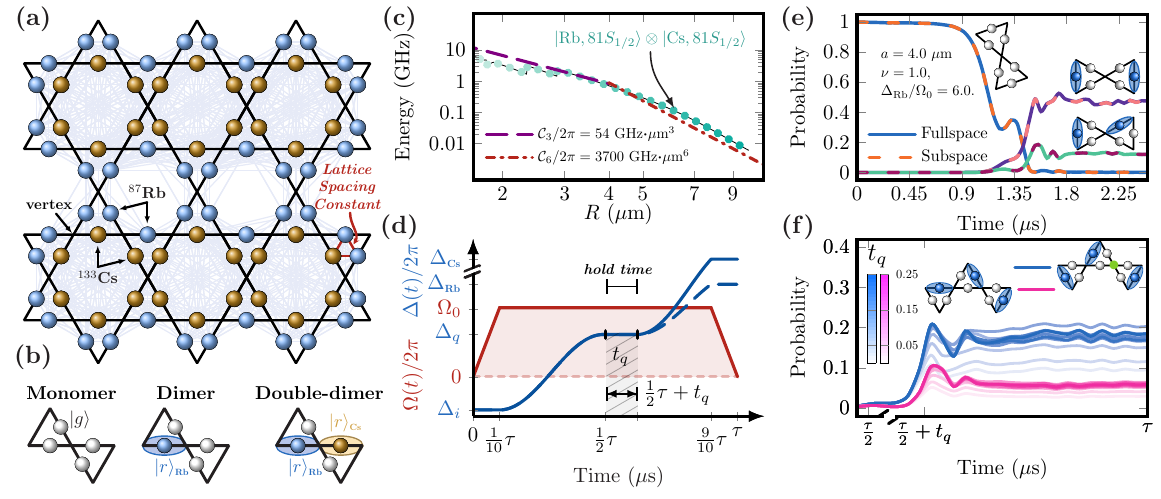}
	\caption{%
		(a) A dual-species array of $^{87}$Rb and $^{133}$Cs atoms shaped as Kagome lattice. It consists of hexagons, each hexagon is surrounded by the triangles consisting of three atoms. The lattice spacing constant $a$ is the distance between any of   two atoms in the corner-sharing triangle. The distance between two atoms located on the two neighboring sides of any hexagon's vertex is $2a$. Rb  and Cs atoms are shown as blue and orange circles, respectively.
		(b) The structure of monomer, dimer, and double-dimer configurations.
		(c) The interaction energy between Rb and Cs atoms both being excited to Rydberg state $|r\rangle=|81S_{1/2},m_j=1/2\rangle$ as a function of interatomic distance $R$, indicating  the dipole-dipole and van der Waals regimes. 
		(d) The time dependence of the effective two-photon Rabi frequency $\Omega_0$ and detuning $\Delta(t)$ from the Rydberg state $|r\rangle$ for adiabatic Rydberg excitation with holding. 
		(e) We compare the resulting probabilities of three different states of considering the full Hilbert space (solid curves) with the truncated subspace with $R_{s}=\frac{3}{2}a$ (dashed curves) for a quantum system of $N=6$ atoms.
		(f) The calculated time dependence of the probability of a system with $N=9$ atoms. The perfect dimer covering state is $|\psi_{\text{PD}}\rangle=|g,r,g,g,r,g,g,r,g\rangle$ (red curves) and the monomer-dimer state is $|\psi_{\text{MD}}\rangle=|r,g,g,r,g,g,g,r,g\rangle$ (blue curves). Here we use the lattice spacing constant $a=4.5$~\si{\mu m} and the detuning $\Delta_{\text{Rb}}/\Omega_{0}=6$. The sweep-hold-sweep protocol  with $\Delta_{q}=0$ is applied for a holding period $t_{q}$ in the range from $0$ to $\frac{1}{10}\tau$~\si{\mu s} with $\nu=2$.
	}
	\label{Figure-QS}
\end{figure*}	


Rydberg quantum simulators can address some of the most challenging and longstanding problems in condensed-matter theory~\cite{lewenstein2007ultracold}. They make it possible to discover and investigate unusual phases of matter that emerge under extreme conditions, such as ultralow temperatures or frustrated interactions. Frustration gives rise to extraordinary quantum features in these exotic phases~\cite{cheng2024emergent, gonzalez2025observation}. One such concept, originally studied in condensed-matter physics, is the quantum spin liquid (QSL)~\cite{savary2016quantum}. This refers to a state in which spins remain in a disordered, liquid-like configuration due to quantum fluctuations, despite strong interactions among them even at absolute zero temperature ($T=0$~\si{K}).  QSLs arise from frustration~\cite{lancaster2023quantum}, where competing interactions prevent spins from settling into an ordered state. Unlike conventional magnetic phases, QSLs do not break spin symmetry and lack long-range magnetic order. The frustration underlying QSLs can naturally stem from geometric structures, where competing interactions induce magnetic disorder. 

The Kagome lattice~\iconKagome is a prime example of frustrated geometric structures. It is a two-dimensional periodic structure of equilateral triangles and hexagons. Because of its geometry, an antiferromagnet on Kagome lattice is highly frustrated, as neighboring spins want to anti-align, but triangles make that impossible to satisfy everywhere at once. After first theoretical analysis of zero-temperature phases of neutral atom arrays on  Kagome lattice~\cite{samajdar2021quantum}, the QSLs in dimer models in Rydberg atom arrays were realized experimentally~\cite{semeghini2021probing}. The direct signatures of topological order and quantum correlations were observed. These  results were later used for dynamical preparation of QSLs in Rydberg atom arrays~\cite{giudici2022dynamical}. Topological QSLs states~\cite{verresen2021prediction} can also be optimized and efficiently prepared in scalable simulations using a time-dependent variational Monte Carlo algorithm~\cite{mauron2025predicting}, or by using approximately symmetric neural quantum states~\cite{vu2025optimizing}. The thermodynamic properties of spin liquid states for atoms on the ruby lattice were studied in Ref.~\cite{WangPRL2025}. A quasi-two-dimensional periodic quantum system, that features a topological $\mathds{Z}_2$ spin liquid as its ground state, was constructed basing on blockade graph automorphisms~\cite{MaierPRX2025}. The sweeping of the parameters of a quantum system was used for generation of QSLs and other exotic states of matter~\cite{sahay2022quantum}. The possibility of finding trimer spin-liquid-like resonating valence bond (RVB) state in finite-sized Rydberg arrays on honeycomb lattice was explored~\cite{kornjavca2023trimer}. The existence of QSLs states on square and triangular geometries was demonstrated with high fidelity by exploiting so-called gadgets to encode the constraints on dimer models~\cite{Zhongda2025Gadget}. QSLs have been realized and investigated also for other quantum simulating platforms~\cite{han2012fractionalized, nath2015hexagonal, eassa2024high, li2025disentangling, bornet2026dirac}. 

Dual-species atomic arrays have emerged in the  experiments of the past few years. Dual-species array of Cs and Rb atoms was first proposed for high-fidelity quantum non-demolition state measurements with low crosstalk~\cite{beterov2015rydberg}. Later, the dual-species Rb-Cs array was realized experimentally~\cite{SinghDual2022}. Parallel implementation of CNOT gates in dual-species arrays was proposed theoretically~\cite{farouk2023parallel}. The dual-species controlled-z gate was experimentally demonstrated~\cite{anand2024Dual}. Dual-species arrays can be also used for syndrome measurements  and error correction~\cite{petrosyan2024fast,miles2026qubit}. The  first dual-species experiment probing the many-body interaction dynamics in 1D using a quantum cellular automata was reported in Ref.~\cite{white2026quantum}. A combination of global driving and static processor layouts in dual-species arrays was introduced to   implement a large class of translation-invariant discrete local dynamics~\cite{cesa2026engineering}.

In this paper we investigate a QSL state of a dual-species array of Rb and Cs neutral atoms, placed on a Kagome lattice. We introduce the dual-species array for realization of QSL to allow controlling the dynamics within this phase using the ratio between the Rydberg detunings of the Cs and Rb atoms as a control knob of the system. In this paper, we aim to build the physical model fo a dual-species array and focus on finding the range of system parameters required for generating the QSL state. We use a sweep-hold-sweep protocol to excite the atoms to Rydberg states and tune the holding time and the detuning from the resonance at Rydberg excitation individually for each atomic species. We identify the boundaries of the QSL state by considering the geometric constraint on Kagome lattice for generating the dimer covering based on the filling fraction, correlation length, mutual information, the topological quantum entanglement entropy using Kitaev-Preskill method, and the topological string operators.

This paper is organized as follows. In section~\ref{Sec: QS}, we describe the quantum system under consideration and the structure of the Kagome lattice. We also explore the properties of monomer/dimer configurations, and study the effect of varying the holding time for unequal detuning from the resonance at Rydberg excitation for Rb and Cs atoms. In section~\ref{Sec:TopologicalZ2}, we numerically investigate the probability of finding QSL states in a Kagome lattice with $N=21$ atoms. We study how the average Rydberg density of the system depends on  the interatomic distance and the ratio of detunings from the resonance at Rydberg excitation for Rb and Cs atoms. In section~\ref{Sec:Correlations}, we focus on the correlations between the system components. This section is divided into two subsections: in subsection \ref{subsec:CorrLength}, we examine the correlations between the pairs of atoms, the decay of correlations, and the maximum correlation length; in subsection~\ref{subsec:MutualInfromation}, we consider the correlations between the system regions by calculating mutual information as a function of size of the subset of the system. We examine whether it follows an area-law or a volume-law. In section~\ref{Sec: TQEE} we use the Kitaev-Preskill method to compute the topological quantum entanglement entropy. In section~\ref{Sec:TopoStrings}, we investigate the topological order by evaluating the diagonal topological string operator on closed paths as a nonlocal observable. The results are summarized in section~\ref{Sec: Conclusion}.


\section{Quantum system\label{Sec: QS}}
We consider a dual-species array of Rb and Cs atoms arranged on a Kagome lattice. Its spatial configuration is shown in Fig.~\ref{Figure-QS}(a). The atoms are placed on the centers of the edges of corner-sharing triangles. The many-body Hamiltonian describing the interaction of ultracold neutral atoms with laser radiation and the interatomic interactions is written as

\begin{equation}
	\begin{split}
		\hat{H}=\frac{\Omega(t)}{2}\sum\limits_{i=1}\left( |g\rangle_i \langle r|_i +\text{h.c.} \right) + \sum\limits_{i=1} \Delta_{i}(t) \hat{n}_i +\sum\limits_{i<j} V_{ij} \hat{n}_i \hat{n}_j.
	\end{split}
	\label{Hamiltonian1}
\end{equation}

The first term in Eq.~(\ref{Hamiltonian1}) is the driving part, which describes laser excitation of  atoms from  the ground state $|g\rangle$ to highly excited Rydberg state $|r\rangle$ by a laser pulse with Rabi frequency $\Omega(t)$. The next two terms are responsible for the interatomic interactions $V_{ij}$ and the time-dependent detuning $\Delta_i(t)$ from the Rydberg state. Here $\hat{n}_i=|r\rangle_i\langle r|$ is a Rydberg density operator for \textit{i}\textsuperscript{th} atom, which can be either rubidium or cesium atom.  The index~$i$ is the number of site on the edge of a Kagome lattice, shown in Fig.~\ref{Figure-QS}(a), where the atom is placed.

The atoms are excited from the ground state ($|g\rangle_{\text{Rb}}=|5S_{1/2},F=2\rangle$ for Rb and $|g\rangle_{\text{Cs}}=|6S_{1/2},F=2\rangle$ for Cs) to a highly excited  Rydberg state $|r\rangle=|81S_{1/2},m_j=1/2\rangle$ using two-photon laser excitation via the intermediate state ($|m\rangle_{\text{Rb}}=|6P_{3/2},F=2,m_F=2\rangle$ for Rb and $|m\rangle_{\text{Cs}}=|7P_{3/2},F=2,m_F=2\rangle$ for Cs) by two laser fields with Rabi frequencies $\Omega_{\text{g}\rightarrow\text{m}}=2\pi \times 40$~\si{MHz} and $\Omega_{\text{m}\rightarrow\text{r}}=2\pi \times 50$~\si{MHz}. This excitation scheme is common in modern experiments with Rydberg quantum simulators.
The laser excitation wavelengths at the first step are $420$~\si{nm} for Rb and $460$~\si{nm} for Cs. At the second step of laser excitation the wavelengths are $1013$~\si{nm} for Rb and $1039$~\si{nm} for Cs. 

Laser excitation of atoms located at sites $i$, $j$ ($i\neq j$) into Rydberg states results in long-range Rydberg interactions, which can be either in dipole-dipole  or van der Waals (vdW) regime, depending on the interatomic distance and the choice of quantum states. Figure~\ref{Figure-QS}(b) illustrates the dependence of the inter-species interaction energy of two atoms on interatomic distance in both regimes~\cite{vsibalic2017arc}. For the range of distances, considered in this work, the interaction is in the vdW regime and can be described as $V_{ij}=\frac{C_{6}}{R^{6}}$. We calculated the  the dispersive vdW coefficients  $C_{6}^{\text{Rb-Cs}}=2\pi \times 3700$~\si{GHz\cdot\mu m^6}, $C_{6}^{\text{Rb-Rb}}=2\pi \times 2550$~\si{GHz\cdot\mu m^6} and $C_{6}^{\text{Cs-Cs}}=2\pi \times 2350$~\si{GHz\cdot\mu m^6}~using Alkali Rydberg Calculator\cite{vsibalic2017arc,farouk2024generation}. In the regime of Rydberg blockade~\cite{JakschCirac2000} only one atom can be excited to the Rydberg state within the range of interatomic distances described by the Rydberg blockade radius $R_{b}$~\cite{beterov2015rydberg,vsibalic2017arc}. This effect is crucial for generating   quantum phases of matter and for implementation of entangling quantum gates.

The maximum value of the time-dependent effective two-photon Rabi frequency $\Omega(t)$ is $\Omega_0=\Omega_{g~m} \Omega_{m~r} / (2\Delta_{m})$, where $\Delta_{m}$ is the detuning from the intermediate state $|m\rangle$. The detuning from the resonance for the transition to Rydberg state $\Delta_{i}(t)$ and the Rabi frequency $\Omega(t)$  are swept during the laser pulse using a quasiadiabatic profile, shown in Fig.~\ref{Figure-QS}(c)]. During the sweeping process from the initial negative value of detuning $\Delta_{i}$ to the positive final value $\Delta_{\text{f}}$, a holding step with constant detuning  $\Delta_q$ is included. This quasiadiabatic regime of laser excitation can be obtained for the following sweeping profile of the detuning:

\begin{equation}
\begin{split}
	&\Delta(t)=\Delta_{i}+\chi_1 \sin^{2}\big( \frac{\alpha_{d_1}}{\tau}(t-\frac{1}{10}\tau)\big) \bigg\vert_{\frac{1}{10} \tau < t \leq \frac{1}{2}\tau},
	\\&
	\Delta(t)=\Delta_{q}+\chi_2 \sin^{2}\big( \frac{\alpha_{d_2}}{\tau} (t-t_{q}+\frac{1}{2} \tau ) \big) \bigg\vert_{(t_q-\frac{1}{2}\tau) < t \leq \frac{9}{10}\tau},
\end{split}\label{Eq-SweepingProfile}
\end{equation}
This profile was introduced independently in~\cite{mc2024towards} and~\cite{farouk2024generation} for the aim of flattening the ramp of the sweeping profile in boundaries of the segment. In Eq.~\ref{Eq-SweepingProfile}, each ramp segment starts and ends with zero slope as desired for smoothing the sweeping while allowing holding the sweeping at $\Delta_{q}$ for a holding time $t_{q}$. The span coefficients $\chi_1=\Delta_{q}-\Delta_{i}$, and $\chi_2=\Delta_{f}-\Delta_{q}$ are the detuning-sweep amplitudes for sweeping segments I and II. While $\alpha_{d_1}$,$\alpha_{d_2}$ are the sweep-rate coefficients for segments I and II and both are used to set the angular rate of the $\sin^2$ argument. Both parameters are fixed by the boundary conditions $\Delta_{i}$, $\Delta_{q}$, $\Delta_{f}$, $\tau$, $t_{q}$ to ensure that the ramp $\dot{\Delta}(t)= 0$ at the end of the sweeping segments i.e. $t=\tau/2$ and $t=9\tau/10$, respectively. This profile describes the sweep-hold-sweep process. The holding starts, when the value of detuning reaches $\Delta_{q}$, and lasts during the interval $t_{q}$. We denote the ratio between detunings of Rb and Cs atoms at the end of the protocol as $\nu=\Delta_{\text{Cs}}/\Delta_{\text{Rb}}$. 

This sweep-hold-sweep profile has been considered earlier in a different style~\cite{semeghini2021probing,lukin2024quantum}. In our implementation of the profile, it includes three following steps,  shown in Fig.~\ref{Figure-QS}(b): I- The sweeping from the initial negative value $\Delta_{i}$ of detuning to the holding value $\Delta_{q}$. This step lasts for time $\frac{1}{2} \tau$. II- The holding step in which both the detuning and the Rabi frequency remain constant in order to drive the atoms out of the equilibrium. This step lasts for $t_{q}:0 \rightarrow \frac{1}{10} \tau$. III- The sweeping is resumed, and the sweeping rate of the detuning is increased. Finally, a positive value of detuning $\Delta_{f}$ is reached. Its value is selected individually for each atomic species. The sweeping rate at the third step depends on the final value of detuning and the holding time $t_{q}$. This profile is designed to push the state of the system out of the equilibrium. The sweeping rate $\tilde{s}(t)$   has two segments of ramps. The first ramp lasts during the interval $t:\frac{1}{10}\tau \rightarrow \frac{1}{2}\tau$, where 
\begin{equation}
	\tilde{s}_{1}(t)=\frac{\chi_1 \alpha_{d_1}}{\tau}\sin \big( \frac{2\alpha_{d_1}}{\tau} ( t-\tau/10) \big) \bigg\vert_{t=\frac{1}{10}\tau}^{t= \frac{1}{2}\tau}.
\end{equation}
Here the detuning profile is the same for all atoms of both species, and the sweeping rate depends on the values of $\Delta_{i}$, and $\Delta_{q}$. The second ramp starts after the end of holding  time and lasts during the time interval $t:(\frac{1}{2}\tau+t_{q})\rightarrow\frac{9}{10}\tau$. The sweeping rate in this segment is defined as
\begin{equation}
	\tilde{s}_2(t)=\frac{\chi_2 \alpha_{d_1}}{\tau}\sin \big( \frac{2\alpha_{d_2}}{\tau} ( t - t_{q} - \tau/2) \big) \bigg\vert_{t=\frac{1}{2}\tau+t_{q}}^{t=\frac{9}{10}\tau}.
\end{equation}
This sweeping rate   depends on the final value of detuning $\Delta_{f}$ and holding value $\Delta_{q}$, and the holding time $t_{q}$, which are individually selected for different species. The maximum value for the sweeping rate in the second segment is achieved for the maximum value of $t_{q}$:
\begin{equation}
	\tilde{s}_{2}^{\text{max}}(t)=\frac{\chi_2 \alpha_{d_2}}{\tau}\sin \big( \frac{2\alpha_{d_2}}{\tau} ( t - 3\tau/5) \big) \bigg\vert_{t=\frac{3}{5}\tau}^{t=\frac{9}{10}\tau}.
\end{equation}

Now we consider how the spatial configuration of Rydberg excitations resemble different phases in a QSL theory. In a Kagome lattice, which is illustrated in Fig.~\ref{Figure-QS}(a),  each vertex of each hexagon has a coordination number of~$4$. It means that it is connected to four neighboring atoms, forming a pattern of corner-sharing triangles on hexagon edges. If the spacing between the sites of the atoms located inside each corner-sharing triangle is $a$, the length of the hexagon edge will be $2a$. Each vertex of the hexagon, which forms a \textit{unit cell} \iconUnitCell around it (here the black circle marks a vertex of a hexagon), is connected to four atoms, which are located at unique positions inside the whole lattice. The structure of Rydberg excitations in a unit cell is illustrated in Fig.~\ref{Figure-QS}(d).  By proper choice of the value of the lattice spacing constant $a$, the atoms, forming a unit cell, can all remain in the ground state $|g\rangle$ after the end of laser excitation. This configuration is denoted as \textit{monomer} due to the absence of Rydberg interactions which are equivalent to dimer bonds in a quantum dimer model. We use a green circle \textcolor{green!80!black}{\scalebox{1.50}{$\bullet$}} to indicate a position of a \iconMonomer  monomer vertex  on a whole lattice. 

When one atom in the unit cell is excited to Rydberg state $|r\rangle$, it will form a configuration, denoted as \textit{dimer}  \iconDimer in a quantum dimer model. The Rydberg blockade gives a strong local exclusion rule and naturally enforces the same hard-core constraints that dimers obey. If two non-adjacent atoms of the unit cell are excited to  Rydberg state, it forms a  \textit{double-dimer} configuration , which can be illustrated as \iconDimerDouble or \iconDimerDoubleB. This is a most frustrated configuration, which is due to weak violation of Rydberg blockade.  The position of a double-dimer vertex in the lattice will be indicated as magenta circle \textcolor{magenta}{\scalebox{1.50}{$\bullet$}}. It can usually appear on the lattice edges and results in frustration in a Kagome lattice due to  competing interactions between the atoms on the corner-sharing triangles. 

In our calculations, the system composed of $N$ dual-species atoms, placed on a Kagome lattice, is initially prepared in the  ground state of all atoms $|g\rangle$. It can be denoted as $|\psi_0\rangle=\otimes_{i=1}^{N} |g\rangle_i=$~\iconMonomer for $N=4$ or \iconFKagomeA for $N=30$ for a whole Kagome lattice configuration [see Fig.~\ref{Figure-QS}(a)]. 

After the laser pulse, the system ends in final state $|\psi_{f}\rangle$. In Fig.~\ref{Figure-QS}(d) Rydberg excitation of a single atom to state  $|r\rangle$ is illustrated by a blue circle for Rb atom and by an orange circle for Cs atom. For example, the final state of four atoms  $|\psi_{f}\rangle=|r,g,g,g\rangle =$ \iconDimer for $N=4$. Here the first atom is a Rb atom, excited to Rydberg state $|r\rangle$, while other atoms (which can be either Rb or Cs) remain in the ground state $|g\rangle$.

For a large quantum system with $N \geq 18$, it is difficult to write the Hamiltonian for all  $2^{N}$ states. We adopt working in a truncated subspace of the blockade space by selecting a subspace radius $R_{s}=\frac{3}{2}a$, which showed excellent agreement with the full Hilbert space calculation as shown in Fig.~\ref{Figure-QS}(e) for a quantum system of $N=6$ atoms. This truncated subspace radius reduces the dimensions of lattice, with filled $N$ sites, to a subspace of dimensions $2^{\frac{2}{3}N}$. This constraint on the interaction Hamiltonian  removes almost all interactions between  non-adjacent triangles on a Kagome lattice and demonstrates strong blockade for  atoms, located on the edges of triangles. Therefore, all possible atomic states for each triangle of the lattice are shown as: 
\begin{equation}
	\begin{split}
		\left\{\iconTriangleA,\iconTriangleB,\iconTriangleC,\iconTriangleD\right\}
	\end{split}\label{Eq: Basis-States}
\end{equation}
Here we made an illustration only for the case of Rydberg excitation of Rb atoms. 

However, the probability of generating a double-dimer state is not completely negligible, as we will show  in Fig.~\ref{Figure-Probability}. The atoms, located on edge of the lattice, can demonstrate a nontrivial dynamics. Therefore, out treatment of the subspaces for heteronuclear array adopts the same strategy, as used in  Bloqade.jl Julia package~\cite{bloqade2023quera}.


We intend to obtain the \iconDimer configuration for all unit cells, which describes a perfect dimer covering on a lattice. Dimer covering (often called tiling) is an arrangement where every vertex of the hexagon is covered by only one dimer. The number of possible dimer coverings increases exponentially with the increase of the lattice size. 

In Fig.~\ref{Figure-QS}(e), we illustrate the effect of variation of a holding time on the dynamics of a quantum system for $N=9$ atoms. We plot the probability to find the system in a perfect dimer covering  state \iconDimerNN, and, alternatively, in a monomer-dimer configuration \iconDimerMonomerNN. The probability is calculated as a function of time $\tau$ for different values of the time holding parameter $t_{q}$ in the range $0\rightarrow\tau/10$ for the lattice spacing constant $a=4.5$~\si{\mu m}, the final value of the detuning $\Delta_{\text{Rb}}/\Omega_0=6$, and $\nu=2$. For the selected interatomic distance, the monomer-dimer configuration is more probable than the perfect dimer covering. 

The increase of $t_{q}$  results in the increased probability for both configurations. However, the  dependence of \iconDimerNN and \iconDimerMonomerNN configurations on $\tau$ is non-trivial, and further optimization of their probabilities should be done using quantum control techniques~\cite{kozenko2026numerically}.


\section{\texorpdfstring{$\mathds{Z}_2$}{Z2}-Quantum Spin Liquids \label{Sec:TopologicalZ2}}

\begin{figure*}[t]
	\centering
	\includegraphics[width=\textwidth]{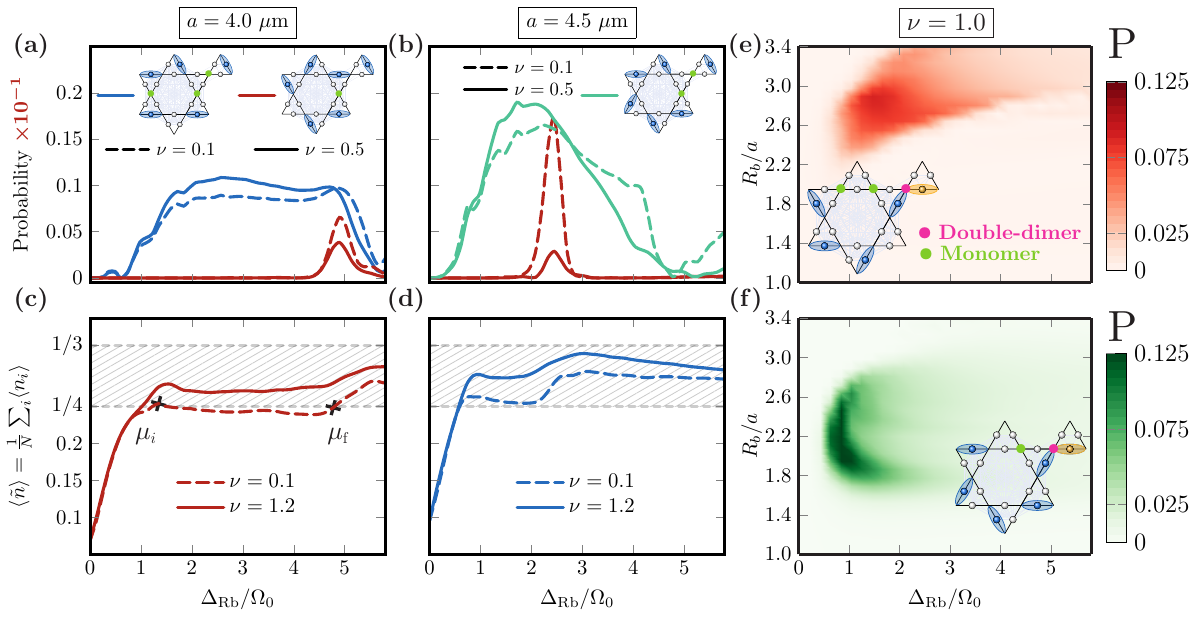}
	\caption{
		(a),(b) The probabilities of exciting a quantum system with $N=21$ atoms on a Kagome lattice to one of three specific configurations of Rydberg excitations, indicated by red, blue and green curves and illustrated on the insets, as a function of the dimensionless parameter $\Delta_{\text{Rb}}/\Omega_0$ for lattice constants (a) $a=4.0$~\si{\mu m}   and (b) $4.5$~\si{\mu m}. The solid and dashed curves illustrate the effect of the ratio $\nu=\Delta_{\text{Cs}}/\Delta_{\text{Rb}}$ on the calculated probabilities. The locations of   monomer vertices are indicated by green circles on the insets. 
		(c),(d) The average Rydberg density $\langle \tilde{n} \rangle=\sum\langle n_i \rangle/N$ for the same system with $N=21$ atoms on a Kagome lattice as a function of the dimensionless parameter $\Delta_{\text{Rb}}/\Omega_0$ for two different values of the ratio $\nu$. The density dependence shown in (c) for $\nu=0.1$ and $a=4.0$\si{\mu m} demonstrates the region $\tilde{n}\leq1/4$ (below the shredded part) which is a geometric constraint for dimer-covering manifold of the Kagome lattice. The geometric constraint for dimer-covering manifold of the Kagome lattice is satisfied in  the range $\mu_{i}\leq \Delta_{\text{Rb}}/\Omega_0
		\leq \mu_{f}$, where $\mu_{i}=1.34$ and $\mu_{f}=4.77$. There is no QSL state for the conditions in (d).
		(e), and (f) The phase diagrams of the probability of generating double-dimer configurations as a function of dimensionless parameters $R_{\text{b}}/a$ and $\Delta_{\text{Rb}}/\Omega_0$. In (e), for short atomic distance $a$, the number of monomer vertices is increased compared to (f).
	}
	\label{Figure-Probability}
\end{figure*}

\begin{figure*}[t]
	\centering
	\includegraphics[width=\textwidth]{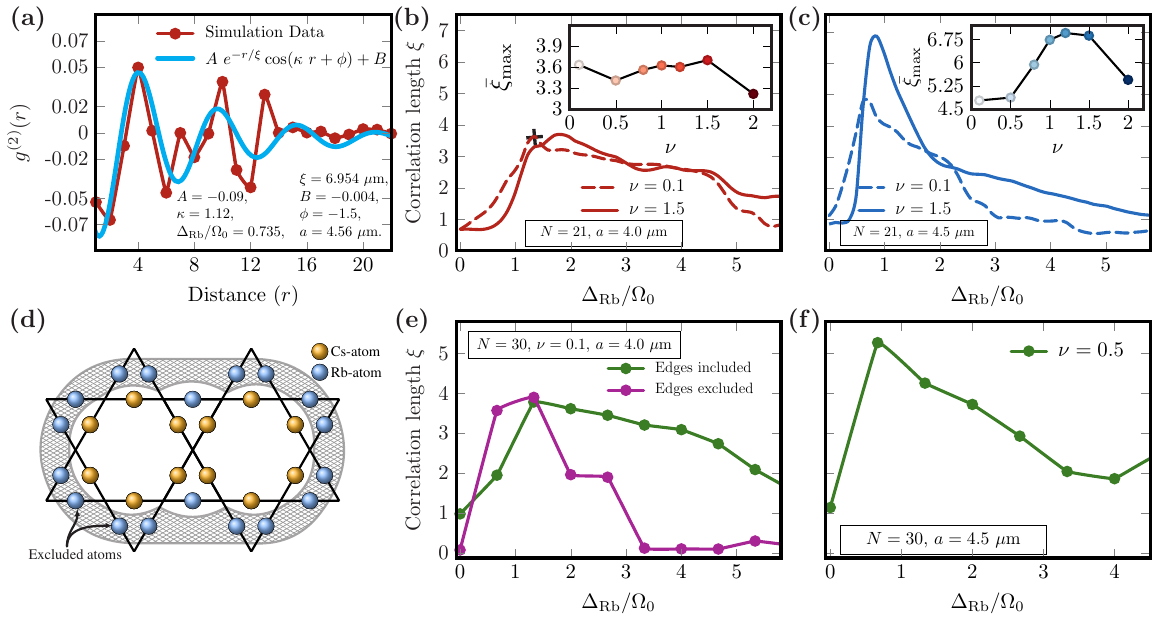}
	\caption{
		\textbf{The calculation of correlation length $\xi$.}
		(a) The density-density correlation function between atoms $g^{(2)}(r)$ as a function of the interatomic distance $r$. It shows the amplitude decay of the	correlations as the distance increases.
		(b) The extracted value of the correlation length $\xi$, calculated as a function of the dimensionless parameter $\Delta_{\mathrm{Rb}}/\Omega_0$ for different configurations of the system. The solid-red and dashed-red  curves represent the cases when $\nu=1.5$  and $\nu=0.1$, respectively, for $a=4.0$~\si{\mu m}, and for the holding time $t_{q}=0.25$. The inset shows the maximum value of the correlation length $\bar{\xi}_{\text{max}}$ as a function of the ratio $\nu$. The correlation length $\bar{\xi}_{\text{max}}$ has maxima at $\nu=0.1$ and $1.5$.
		(c) The same as in (b), but for $a=4.5$~\si{\mu m}.
		(d) Schematic representation of the positions of the edge atoms, which were removed to reduce  the subspace for the calculations of the correlation length $\tilde{\xi}$.
		(e) The correlation length $\xi$ of a system with $N=30$ atoms. The green curve shows the case where all atoms are included in the calculation of $\xi/a$, yielding $\bar{\xi}_{\text{max}}/a \simeq 4$. The violet curve represents the correlation length $\tilde{\xi}/a$ calculated after removing the atoms from the edges of the array [the shaded region in (d)]. The maximum values obtained in both calculations coincide with the left-hand bound of the QSL state $\mu_{i}$ [Fig.~\ref{Figure-Probability}(c)].
		(f) The same as in (e), but with $a = 4.5$~\si{\mu m} and $\nu = 0.5$.
	}
	\label{Figure-CorrLength}
\end{figure*}

QSL states emerge due to the frustration, where the vdW interactions prevent spins from settling into an ordered state, or a commensurate phase of matter~\cite{verresen2021prediction,semeghini2021probing}. In Figs.~\ref{Figure-Probability}(a),(b), the probabilities of generation of different monomer-dimer configurations are illustrated as a function of the dimensionless parameter $\Delta_{\text{Rb}}/\Omega_0$. The calculations are made for a quantum system with $N=21$ atoms and with two different  lattice spacing constants $a=4.0$~\si{\mu m} in Fig.~\ref{Figure-Probability}(a) and $a=4.5$~\si{\mu m} in Fig.~\ref{Figure-Probability}(b), each time for two different values of the ratio $\nu$. From Fig.~\ref{Figure-QS}(e) we clearly see that with the increase of $\nu$ the probability to obtain the corresponding configuration behaves in a sinusoidal manner.  Figures~\ref{Figure-Probability}(a, b) show three monomer-dimer configurations \iconEnKagomeMDimerA, \iconEnKagomeMMDimerA, and \iconEnKagomeMMMDimerA. The number of Rydberg excitations in each configuration is increased for larger interatomic distances, as it is clearly seen for $\nu=0.5$. The evolution of the mean value of a Rydberg density $\langle \tilde{n} \rangle=\sum_i \langle n_{i} \rangle/N$ as a function of the dimensionless parameter $\Delta_{\text{Rb}}/\Omega_0$ for these two cases is shown in Figs.~\ref{Figure-Probability}(c),(d). In Fig.~\ref{Figure-Probability}(c), we find the bulk density $\tilde{n}<1/4$ for $a=4$~\si{\mu m} and $\nu=0.1$, which occurs within the bounded region $\mu_{i}\leq \Delta_{\text{Rb}}/\Omega_0 \leq \mu_{f}$. Here $\mu_{i}=1.34$ and $\mu_{f}=4.77$. In Fig.~\ref{Figure-CorrLength}, the maximum of the correlation length $\xi$ is reached at $\Delta_{\text{Rb}}$, which coincides with $\mu_{i}$. It satisfies the condition of a filling fraction corresponding to the QSL state. This means that in this case the resulting state does not have any spatial arrangement. In the experiment, a different configuration, resembling Figs.~\ref{Figure-Probability}(a),(b), can appear after each repetition of laser excitation.

For $\nu>0.1$ the bulk density $\tilde{n}$ remains bounded between values $1/4 < \tilde{n} \leq 1/3$ for any value $\Delta_{\text{Rb}}/\Omega_0$ greater than the value of critical detuning at which the . This marks the first maximum in the evolution of $\tilde{n}$.

In Fig.~\ref{Figure-Probability}(e-f), we show the phase diagram of the probability of finding a double-dimer configuration as a function of dimensionless parameters $R_{b}/a$, and $\Delta_{\text{Rb}}/\Omega_{0}$ when $\nu=1$, for both phases. It can be  noticed that the probability of finding a double-dimer configuration has a maximum around $\Delta_{\text{Rb}}/\Omega_0\simeq1$, where the position of this double-dimer vertex is mostly affected by being at edge of the lattice [see the magenta dot marking the position of the vertex]. The probability drops drastically for larger values of $\Delta_{\text{Rb}}/\Omega_0$. For a shorter lattice spacing constant $R_{b}/a > 2.4$, the number of monomers is increased. 


\section{Correlations\label{Sec:Correlations}}

Studying the correlations between the connected systems or subsystems aims to quantify the information, that can be acquired from one system(subsystem), about the other. There is a common way of studying the correlation between two points, which are separated by a distance $r$.  In subsection~\ref{subsec:CorrLength} we investigate the decay of the correlation with increase of interatomic distance. Another approach is calculating or measuring the correlations between different connected regions of the system, and studying the relation between the correlation and the size of that region. The original interest in this relation came from the discovery that the amount of entropy of black holes scales as the area of surfaces at the event horizon (area-law)~\cite{hooft1985quantum}. For the systems of ultracold  atoms, even at zero temperature,  all correlations arise due to entanglement, which is characterized by entropy (entanglement entropy). In subsection~\ref{subsec:MutualInfromation}, we study the mutual information between the regions or subsets of the  quantum system, which coincides with the entanglement entropy at ultra-low (almost zero) temperatures.


\subsection{Correlation length \texorpdfstring{$\xi$}{xi}\label{subsec:CorrLength}}

The strength of any phase of matter is characterized by the correlation length $\xi$, which is a characteristic distance that defines the range of influence of the atom on the other atoms. To find it, we calculate a second-order correlation function: 
\begin{equation}\label{Eq-CorrFun}
	g^{(2)}(r)=\sum\limits_{i} \bigl( \langle n_{i} n_{i+r}\rangle- \langle n_{i} \rangle \langle n_{i+r} \rangle \bigl)\frac{1}{N_{r}}	
\end{equation}
where $\langle n_{i} \rangle$ is the Rydberg density of atoms at site $i$ and  $\langle n_{i+r} \rangle$ is the Rydberg density of the other site, which is separated from the site $i$ by distance $r$. Here $N_{r}$ is the normalization constant. Its value is given by the number of pairs of sites or atoms,  which are separated by distance $r$. The correlation function $g^{(2)}(r)$  describes the spatial relationship between different atoms, separated by distance $r$. At zero distance, the correlation function becomes a Mandel $Q$ parameter. In Fig.~\ref{Figure-CorrLength}(a), we show the evolution of the correlation function $g^{(2)}(r)$ for $N=21$ atoms, as a function of  distance $r$ in the pairs of atoms. It shows that the correlation decays to zero with increase of the distance $r$. The correlation length $\xi$ can be extracted from the correlation function by fitting the data for different parametrization~\cite{lienhard2018observing,keesling2019quantum, sahay2022quantum, zhang2025probing}. In our calculations we considered the following approximation of the correlation function:
\begin{equation}
	\tilde{g}^{(2)}(r)=A e^{-r/\xi} \cos(\kappa r+\phi)+B.
\end{equation}
The chosen driving parameter $\nu=0.1$ and  the lattice constant $a=4$~\si{\mu m}  are suitable for obtaining the QSL state, as we previously discussed. The value $\xi/a=3.6$ reaches its maximum   at $\Delta_{\text{Rb}}/\Omega_{0}=1.34$, which correspond to  the left-hand boundary $\mu_{i}$ in Fig.~\ref{Figure-Probability}(c). The observed correspondence between the maximum value of the correlation length $\xi$ and the boundaries of the QSL states is in agreement with the previous work of Verresen et al.~\cite{verresen2021prediction}. As $\xi/a$ decreases, the relation $\Delta_{\text{Rb}}/\Omega_{0}$ grows. For the same interatomic distance, when $\nu=1.5$, the value  $\xi/a$ reaches its maximum   $\bar{\xi}_{\text{max}}/a=3.6$. The evolution of the maximum value of the correlation length $\bar{\xi}_{\text{max}}/a$ as a function of ratio $\nu$ is shown on the inset of Fig.~\ref{Figure-CorrLength}(b). By increasing the interatomic distance to $a=4.5$~\si{\mu m}, as shown in Fig.~\ref{Figure-CorrLength}(c), the quantum system transforms to a different phase of the matter. The correlation length grows sharply to $\xi/a=4.8$ for $\nu=0.1$ and to $\xi/a=6.9$ for $\nu=1.5$. The  inset of Fig.~\ref{Figure-CorrLength}(c) demonstrates the maximum value $\xi_{\text{max}}/a=6.95$ at $\nu=1.2$. 


\subsection{The mutual information \label{subsec:MutualInfromation}}
\begin{figure*}[t!]
	\centering
	\includegraphics[width=\textwidth]{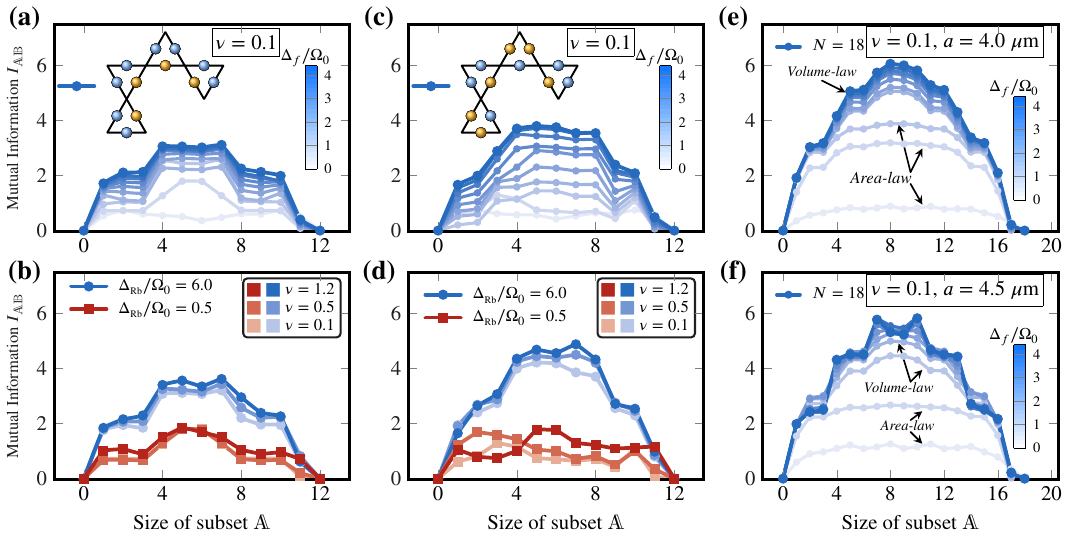}
	\caption{
		\textbf{Mutual information $I_{\mathds{A}\mathds{B}}$.} We divided a system of $N=12$ and $18$ atoms placed on a Kagome lattice to two subsets $\mathds{A}$ and $\mathds{B}$ s.t. $\mathds{A} \cap \mathds{B}=\emptyset$. We show here the evolution of the mutual information $I_{\mathds{A}\mathds{B}}$, as the size of the subset $\mathds{A}$ increases from $0$ to $N$.
		(a),(b) The mutual information of $N=12$ atoms when $1/3$ of all atoms are Cs atoms, distributed as shown in the scheme in the inset of (a). In (a), we show the evolution of the mutual information with the system size, as the value of $\Delta_{\text{Rb}}/\Omega_{0}$ increases, with $\nu=0.1$. In (b), we show the evolution for three different values of $\nu$, as indicated.
		(c),(d) The same as in (a) and (b) but for the case when $1/2$ of all atoms are Cs atoms, with positions shown in the inset on (c).
		(e),(f) The case of $N=18$ atoms with $\nu=0.1$ for two different lattice spacing constants $a=4.0$~\si{\mu m} in (e) and $a=4.5$~\si{\mu m} in (d). We can see here a clear case of the Area-law  and Volume-law evolutions of the mutual information [see the main text]. The distribution of Rb and Cs atoms here is the same as the inset of (a).
	}
	\label{Figure-MutualInfo}
\end{figure*}

The locality of interactions\footnote{It means that any atom in the system is interacting with a finite number of neighboring atoms over a short distance} in a quantum many-body system does not only inherit a decay in correlations, as shown in subsection~\ref{subsec:CorrLength}.  It is also reflected in the scaling laws of the topological entanglement entropy of the ground state~\cite{eisert2010colloquium, islam2015measuring, bluvstein2022quantum, farouk2023scalable}. Due to the entanglement, the quantum systems at zero temperature have nonzero entropy of a subspace of the system. The properties of entanglement entropy were studied intensively for atoms on a Kagome lattice at the boundaries of QSLs state~\cite{verresen2021prediction}. We will quantify the amount of entropy using von Neumann entropy $S(\varrho_{\mathds{A}})$, which is a special case of Renyi entropy $S_{n}(\varrho)=\ln \text{Tr}\varrho^n /(1-n)$ when $n\rightarrow1$~\cite{Horodecki2009}. For a subregion or subset of the quantum system $\mathds{A}$ the entropy is given by
\begin{equation}
	S(\varrho_{\mathds{A}})=-\text{Tr} \left( \varrho_{\mathds{A}} \ln \varrho_{\mathds{A}} \right),\label{Eq-VNE}
\end{equation}
where $\varrho_{\mathds{A}}=\text{Tr}_{\mathds{B}} \varrho$ is the reduced density operator of the subsystem $\mathds{A}$ and $\mathds{B}=\mathds{A}^{\text{c}}$ is the complement of the subsystem $\mathds{A}$. The mutual information reflects the total amount of correlations between two systems. It can show interesting scaling properties with respect to the subsystem size $\mathds{A}$. It is defined as:
\begin{equation}
	I_{\mathds{AB}}=S_{\mathds{A}}+S_{\mathds{B}}-S_{\mathds{AB}},
	\label{Eq-MInfo}
\end{equation}

\begin{figure*}[t]
	\centering
	\includegraphics[width=\textwidth]{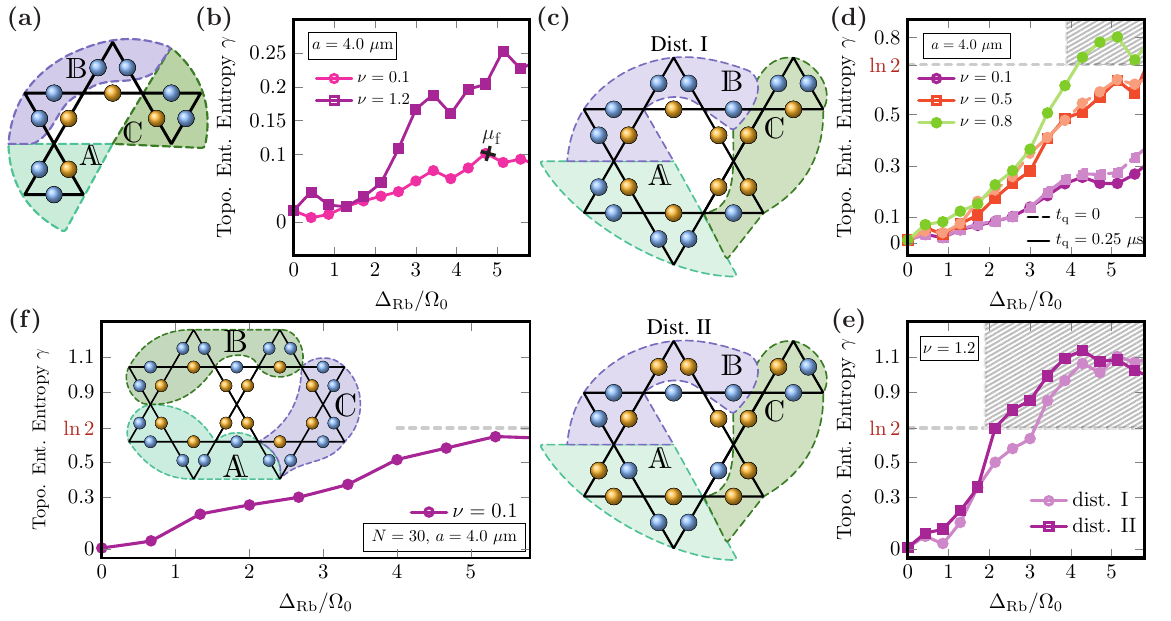}
	\caption{\textbf{The topological entanglement entropy $-\gamma$}. 
		(a) For an array of $N=12$ atoms, we selected three non-intersecting subsets of atoms $\mathds{A}$, $\mathds{B}$ and $\mathds{C}$. There are two atoms, which are not included in any of the subsets. The subsets $\mathds{A}$ and $\mathds{B}$ ($\mathds{B}$ and $\mathds{C}$) have shared-edges, while $\mathds{A}$ and $\mathds{C}$ does not have any shared edges.
		(b) For lattice spacing constant $a=4.0$~\si{\mu m} the  Kitaev-Preskill topological entanglement entropy is calculated for the subsets, shown in (a), as a function of the dimensionless parameter $\Delta_{\text{Rb}}/\Omega_{0}$ for two different values of $\nu=0.1$ in (b) and $\nu=1.2$ in (c). In (b), we note that $0<\gamma\leq 0.1$, and we can see that there are two peaks for $\gamma$ at $\Delta_{\text{Rb}}/\Omega_{0}=3.42$ and $\Delta_{\text{Rb}}/\Omega_{0}=4.7$. The maximum value of $\gamma$ occurs when $\Delta_{\text{Rb}}/\Omega_{0}$=$\mu_{f}$, which corresponds to the bound of the QSL states as shown in Fig.~\ref{Figure-Probability}(c). In (c), $0<\gamma\leq 0.25$ also reaches the first maximum, as in (b), when $\Delta_{\text{Rb}}/\Omega_{0}=3.42$, and the second maximum $\Delta_{\text{Rb}}/\Omega_{0}=5.12$.
		(c) For the array of $N=21$ atoms we have selected three non-intersecting subsets, using the same strategy as in (a), but here there are three unselected atoms, and the subsets $\mathds{A}$ and $\mathds{C}$ have shared edges.
		(d),(e) We consider the system described in (d) with $N=21$ atoms. In (e), we illustrate the dependence of $\gamma$  on the holding time parameter $t_{q}$. The solid and dashed curves correspond to $t_{q}=0$  and $t_{q}=0.25$~\si{\mu s}, respectively. In (f), we illustrate the evolution of $\gamma$ for two different spatial distributions of Cs atoms in the array of $N=21$ atoms. Distribution~I corresponds to the Cs atoms, located as in (d). Distribution~II corresponds to the same configuration, as in Fig.~\ref{Figure-MutualInfo}(c).
		(f) The evolution of TQEE vs. $\Delta_{\text{Rb}}/\Omega_{0}$ for $N=30$ with $\nu=0.1$ and $a=4.0$~\si{\mu m}. In the inset scheme of the subsets $\mathds{A}$, $\mathds{B}$ and $\mathds{C}$.
	}
	\label{Figure-TEE}
\end{figure*}


In Fig.\ref{Figure-MutualInfo}, we show the evolution of mutual information $I_{\mathds{AB}}$ for different system parameters as a function of  size of the subset $\mathds{A}$. In Figs.~\ref{Figure-MutualInfo}(a-d) we consider a system of $N=12$ atoms with the lattice constant $a=4.0$~\si{\mu m} for two different spatial configurations of the dual-species array. In Figs.~\ref{Figure-MutualInfo}(a, b) Cs atoms are located in the inner shell of the array. They consist $1/3$ of the atoms in the lattice, while in Figs.~\ref{Figure-MutualInfo}(c, d) Cs atoms are located both in the inner and outer shells of the triangles of the lattice. In the latter case the number of Cs atoms is $1/2$ of the whole number of atoms. 

From the calculations it is clear that for the same lattice and laser excitation parameters the amount of mutual information slightly increases due to the change of the lattice spatial configuration and structure. In Figs.~\ref{Figure-MutualInfo}(a, c) we show the case with $\nu=0.1$. The color of $I_{\mathds{AB}}$ curve indicates the change of the value of $\Delta_{\text{Rb}}/\Omega_{0}$. For smaller values of $\Delta_{\text{Rb}}/\Omega_{0}\simeq0.5$, we see that $I_{\mathds{AB}}$ almost does not change, as the size of $\mathds{A}$ increases, following the Area-law. But for larger values of $\Delta_{\text{Rb}}/\Omega_{0}$, $I_{\mathds{AB}}$ increases and becomes nearly constant when the size reaches $\mathds{A}\in ]\frac{N}{2}-2,\frac{N}{2}+2]$. 

The same behavior can be observed for larger arrays. In Figs.~\ref{Figure-MutualInfo}(e, f) we consider a large array of $N=18$ atoms and $\nu=0.1$. For $\Delta_{\text{Rb}}/\Omega_{0}\leq 1$, the amount of mutual information almost fixed or increases slowly as the size of the subset increases which means that the system follows the area-law, and the mutual information sharply increases, as the size of subset $\mathds{A}$ is around the half of the array $N/2$. In Fig.~\ref{Figure-MutualInfo}(f), we set the lattice spacing constant $a=4.5$~\si{\mu m}, and the system demonstrates almost similar behavior, but the value of $I_{\mathds{AB}}$ drops slightly around the half of the array. 

In Figs.~\ref{Figure-MutualInfo}(b,d), we see the effect of larger values of $\nu$. Regardless the structure of the array, and of the value of $\nu$, for small values of $\Delta_{\text{Rb}}/\Omega_{0}\leq 1$, the system almost follows the area-law. For larger values of $\Delta_{\text{Rb}}/\Omega_{0}$, the system follows the volume-law.


\section{Kitaev-Preskill Topological Quantum Entanglement Entropy~\label{Sec: TQEE}}

For further investigation of the long-range entanglement of the dynamically prepared state, we calculate the topological quantum entanglement entropy (TQEE) $-\gamma$ over a subset $\mathds{D}$ of the system. This subset is divided into three non-intersecting subsets, such as $\mathds{D}= \mathds{A} \cup \mathds{B} \cup \mathds{C}$. We use the Kitaev-Preskill formula~\cite{kitaev2006topological}:
\begin{equation}
	-\gamma=S_{\mathds{A}}+S_{\mathds{B}}+S_{\mathds{C}}-S_{\mathds{AB}}-S_{\mathds{AC}}-S_{\mathds{BC}}+S_{\mathds{ABC}}.
\end{equation}
where $S_{\mathds{A}}$ is the von Neumann entropy, defined in Eq.~\ref{Eq-VNE} for the subset $\mathds{A}$, and $S_{\mathds{AB}}=S_{\mathds{A}\cup\mathds{B}}$. This formula is a universal characterization of a many-particle quantum entanglement in the ground state of a topologically ordered two-dimensional array~\cite{giudici2022dynamical, mauron2025predicting}. For the most known 2D topologically ordered phases, the scaling is  $\gamma=\log(\mathcal{D})$, where $\mathcal{D} \geq 1$ is a total quantum dimension. Since the quantum dimension $\mathcal{D} \geq 1$, we obtain $\gamma \geq 0$, i.e. the value of TQEE can not be negative. We find $\gamma=0$ if $\mathcal{D}=1$, which is a trivial topologically ordered phase (i.e. no anyons). The case when $0<\gamma\leq \log(2)=0.693$ is the simplest non-trivial topologically ordered phase. The value $\gamma=\log(2)$ represents the $\mathds{Z}_{2}$ toric code topological ordered phase. Quantum phases with $\gamma>0$ are robust against local perturbations. They host anyons, which makes them useful for fault-tolerant quantum computations.

Figure~\ref{Figure-TEE} illustrates  the TQEE for $N=12$ and $N=21$ for different values of the system parameters. In Fig.~\ref{Figure-TEE}(a), we show a scheme representing the subsets $\mathds{A}$, $\mathds{B}$ and $\mathds{C}$ where $\mathds{D}=\mathds{A}\cup\mathds{B}\cup\mathds{C} \subset N$. We selected the subsets to be non-intersecting and the pairs of subsets ($\mathds{A}$ and $\mathds{B}$), and ($\mathds{B}$ and $\mathds{C}$) have shared edges while the subsets $\mathds{A}$ and $\mathds{C}$ do not have them. In Figs.~\ref{Figure-TEE}(b-c), we see the evolution of $\gamma$ as a function of $\Delta_{\text{Rb}}/\Omega_{0}$ for (b) $\nu=0.1$  and for (c) $\nu=1.2$. The evolution of $\gamma$ by increasing the value of $\Delta_{\text{Rb}}/\Omega_{0}$ for $\nu=0.1$ is steady, and has two peaks at $\Delta_{\text{Rb}}/\Omega_{0}=3.42$ and $\Delta_{\text{Rb}}/\Omega_{0}=4.7$. This corresponds to the value $\mu_{f}$ [see Fig.~\ref{Figure-Probability}(c)], which determins a bound of a QSL state. By considering larger value of $\nu=1.2$, in Fig.~\ref{Figure-TEE}(c), the value $\gamma$ increases, and also has one of its peaks, which coincides with the case of $\nu=0.1$ in Fig.~\ref{Figure-TEE}(b). In Figs.~\ref{Figure-TEE}(e, f) we consider a larger array of $N=21$ atoms and a different way of constructing the subsets $\mathds{A}$, $\mathds{B}$ and $\mathds{C}$, which is illustrated in Fig.~\ref{Figure-TEE}(d). 

In  Fig.~\ref{Figure-TEE}(d), the subsets $\mathds{A}$ and $\mathds{C}$ have a shared edge, unlike the previous case in Figs.~\ref{Figure-TEE}(b,c). In Fig.~\ref{Figure-TEE}(b), we select different values of $\nu$ and of the holding time parameter $t_{q}$. The  solid-violet curve for $t_{q}=0$ ($t_{q}=0.25$~\si{\mu s}) and the dashed curve for $\nu=0.1$ show that larger value of $t_{q}$ slightly reduce the value of $\gamma$ for $\Delta_{\text{Rb}}/\Omega_{0}>4.0$.

However, for $\nu>0.1$, the value of $\gamma$ is robust  to the change in $t_{q}$. Also, we can see that the entropy remains in the range $0 < \gamma \leq \log(2)$ for moderate values of $\nu$ and $\Delta_{\text{Rb}}/\Omega_{0}$. In Fig.~\ref{Figure-TEE}(f), we plot the evolution of $\gamma$ for two different spatial distributions of Cs atoms in the array. Distribution~I is the case, sketched in Fig.~\ref{Figure-TEE}(d). Here Cs atoms are located in the inner shell of the Kagome array and exhibit $\sim 42\%$ of the atoms in the array. Distribution~II is the case when Cs atoms are located both in the inner and outer shells of the Kagome lattice, as shown in Fig.~\ref{Figure-MutualInfo}(c), and exhibit $
\sim 61\%$ of all atoms. We can see that the values of $\gamma$ are almost the same for both distributions in the range $\Delta_{\text{Rb}}/\Omega_{0}<\sqrt{3}$. For larger values  $\gamma$  sharply increases to values larger than $\log(2)$, which is the optimum value for toric code implementation. For the configuration with Distribution~I the increase to $\log(2)$ is a bit slower.

In Fig.~\ref{Figure-TEE}(f), we show the evolution of the TQEE for $N=30$ atoms, $\nu = 0.1$ and $a = 4.0$~\si{\mu m}. The TQEE values remain below the upper bound of the toric code topological ordered phase, which is consistent with the results for $N=21$, presented in Fig.~\ref{Figure-TEE}(d) for the same value of $\nu$.  In the inset of  Fig.~\ref{Figure-TEE}(f), we illustrate the subsets $\mathds{A}$, $\mathds{B}$, and $\mathds{C}$.

\section{Topological String Operators\label{Sec:TopoStrings}}
\begin{figure*}[t]
	\centering
	\includegraphics[width=\textwidth]{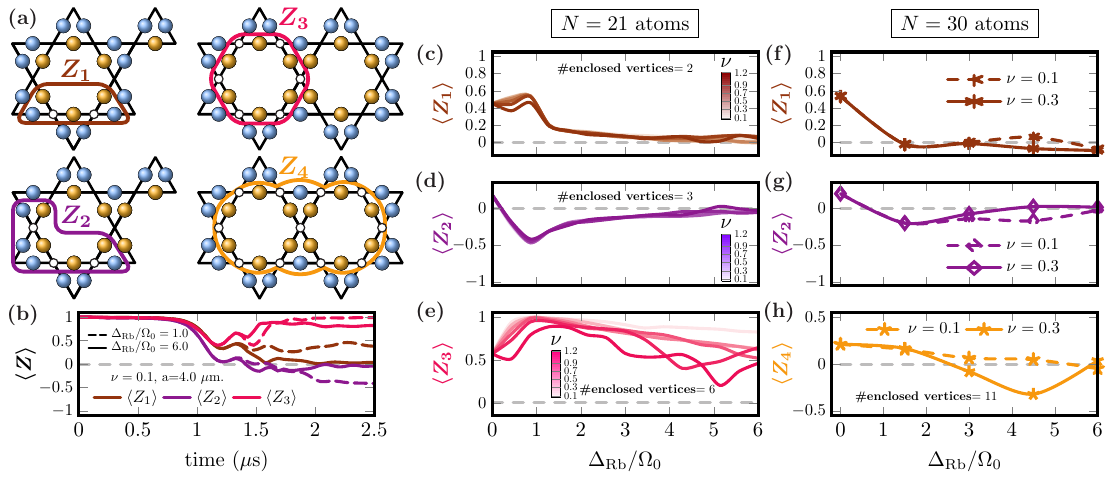}
	\caption{\textbf{Topological diagonal string operators}. 
		(a)Schemes of closed loops for the diagonal string operator $\langle Z \rangle$ on the Kagome lattice: $Z_1$ is a trapezoidal loop with 2 enclosed vertices, $Z_2$ is a L-shaped loop with 3 enclosed vertices, $Z_3$ is a hexagonal loop of the inner shell of the Kagome lattice with 6 enclosed vertices, and $Z_4$ is a 11-vertices loop.
		(b) The evolution of $\langle\hat{Z}_1\rangle$, $\langle\hat{Z}_2\rangle$, and
		$\langle\hat{Z}_3\rangle$ as a function of the profile implementation time during the quasi-adiabatic sweep, for $\nu=0.1$ and $a=4.0$~\si{\mu m}, comparing the evolution when $\Delta_{\text{Rb}}/\Omega_0=1.0$ (dashed-curves) and $\Delta_{\text{Rb}}/\Omega_0=6.0$, (solid-curves).
		(c-e) The evolution of the different closed loops $\langle\hat{Z}_i\rangle$ ($i=1,2,3$) as a function of $\Delta_{\text{Rb}}/\Omega_0$ for six values of the dual-species detuning ratio $\nu=\Delta_{\text{Cs}}/\Delta_{\text{Rb}} \in [0.1,1.2]$.
		(f-h) $\langle\hat{Z}_i\rangle$ ($i=1,2,4$), now evaluated for a system of $N=30$ atoms, shown for two representative values of the dual-species detuning ratio, $\nu=0.1$ (dashed-curves) and $\nu=0.3$ (solid-curves).
	}
	\label{Figure-TSOP-Z}
\end{figure*}
While the average Rydberg density and the correlation length quantify local and short-range properties of the prepared state, they cannot by themselves reveal the presence of topological order, since topological order is, by definition, not accessible to any local measurement~\cite{wen2017colloquium, kitaev2003fault}. To probe the onset of a $\mathds{Z}_2$ quantum spin liquid directly, it is necessary to evaluate nonlocal observables known as topological string operators, defined along closed as in Fig.~\ref{Figure-TSOP-Z}(a), threading the Kagome lattice~\cite{verresen2021prediction,semeghini2021probing, mauron2025predicting, kornjavca2023trimer}. We consider two complementary string operators, a diagonal operator $\hat{Z}=\prod_{i\in S}\hat{\sigma}^{z}_{i}$, with $\hat{\sigma}^{z}_{i}=1-2\hat{n}_{i}$, that measures the parity of Rydberg excitations along a string $S$ of atoms perpendicular to the bonds of the Kagome lattice. In terms of the basis states defined in Eq.~(\ref{Eq: Basis-States}), it can be represented as~\cite{semeghini2021probing}
\begin{equation}
\hat{Z}= \iconTriangleStringZ : 
\left\{\hskip0.5cm
\begin{split}
	\iconTriangleA \tikz\draw[-latex,color=black, line width=1.5 pt] (0,0) to (1.10,0); \iconTriangleA \\
	\iconTriangleB \tikz\draw[-latex,color=black, line width=1.5 pt] (0,0) to (1.10,0); \iconTriangleB\\
	\iconTriangleC \tikz\draw[-latex,color=black, line width=1.5 pt] (0,0) to (0.7,0); -\iconTriangleC\\
	\iconTriangleD \tikz\draw[-latex,color=black, line width=1.5 pt] (0,0) to (0.65,0); -\iconTriangleD\\
\end{split}
\right.,\label{Eq: DiagonalOperator}
\end{equation}
For a perfect dimer covering of a Kagome lattice, $\langle \hat{Z} \rangle = (-1)^{\#\text{enclosed vertices}}$, so that closed $Z$-loops directly diagnose the dimer character of the state.


In Fig.~\ref{Figure-TSOP-Z}, we evaluate $\langle \hat{Z}\rangle$ on closed loops of different sizes, for our dual-species arrays of $N=21$ and $N=30$ atoms, and use their combined behavior across $\Delta_{\text{Rb}}/\Omega_{0}$ to identify the parameter range consistent with a QSL phase. In Fig.~\ref{Figure-TSOP-Z}(a), we show different closed loops for the diagonal string operators: $Z_1$, a trapezoidal loop with 2 enclosed vertices; $Z_2$, an L-shaped loop with 3 enclosed vertices; $Z_3$, the hexagonal boundary loop with 6 enclosed vertices; and $Z_4$ larger loop with 11 enclosed vertices. In Fig.~\ref{Figure-TSOP-Z}(b), we show the evolution of $\langle\hat{Z}_1\rangle$, $\langle\hat{Z}_2\rangle$, and $\langle\hat{Z}_3\rangle$ as a function of the time over the course of the sweeping profile when $\nu=0.1$, $a=4.0$~\si{\mu m} and $t_{q}=0.25$~\si{\mu s}. We compare the evolution for two values of the final detuning $\Delta_{\text{Rb}}/\Omega_0=1.0$, (dashed-curves)  with $\Delta_{\text{Rb}}/\Omega_0=6.0$, (solid-curves). All three operators
begin at $\langle\hat{Z}\rangle=1$, as expected for the unexcited
initial state $|g\rangle^{\otimes N}$. As the detuning sweeping process starts with reaching the maximum value of the Rabi frequency at
$t\geq \frac{1}{10}\tau=0.5$~\si{\mu s}, the three strings fall together, before
separating according to loop parity.  $\langle\hat{Z}_1\rangle$ interpolates between
the two, settling near $0.4$--$0.6$. $\langle\hat{Z}_2\rangle$ is pulled toward negative values reaching $\approx-0.4$ for $\Delta_{\text{Rb}}/\Omega_0=6.0$, consistent with its odd vertex count. $\langle\hat{Z}_3\rangle$ recovers most strongly toward positive values $\approx0.8$--$0.9$ at the end of the sweeping process $t=2.5$~\si{\mu s}), consistent with its even number of enclosed vertices.
 
In Fig.~\ref{Figure-TSOP-Z}(c-e), we show the same three operators evaluated at the end of the sweep as a function of the final detuning $\Delta_{\text{Rb}}/\Omega_0$, for
six values of the dual-species detuning ratio $\nu\in[0.1,1.2]$.
$\langle\hat{Z}_1\rangle$ (panel c) rises to a modest local maximum
($\approx0.5$) near $\Delta_{\text{Rb}}/\Omega_0\approx1$ before decaying
toward zero as $\Delta_{\text{Rb}}/\Omega_0 \geq 4.0$, and shows
comparatively weak dependence on $\nu$. $\langle\hat{Z}_2\rangle$ (panel d)
instead develops negative parity, reaching $\approx-0.5$ near
$\Delta_{\text{Rb}}/\Omega_0\approx1$--$1.5$, in qualitative agreement with
the expected odd-parity sign for a 3-vertex loop, before relaxing back
toward zero at larger value of detuning. $\langle\hat{Z}_3\rangle$ (panel e), whose loop lies entirely on Cs atoms, exhibits the richest structure of the three string operators. Since its loop lies entirely on Cs atoms, all six curves coincide at $\Delta_{\text{Rb}}=0$, regardless of $\nu$ before growing to a broad plateau $\langle\hat{Z}_3\rangle\approx0.9$--$1.0$ for
$\Delta_{\text{Rb}}/\Omega_0\approx1$--$2$, consistent with the even-parity
expectation for this 6-vertex loop, before the curves fan out at larger
detuning in a pronounced oscillatory decay. This is both the largest-magnitude and the most $\nu$-sensitive of the three signals, suggesting that the hexagonal loop is
the most diagnostic probe of how the dual-species detuning ratio governs the
crossover into the dimer-covering regime.

In Fig.~\ref{Figure-TSOP-Z}(f-h), we show the string loops for the $N=30$ array, evaluating $\langle\hat{Z}_1\rangle$, $\langle\hat{Z}_2\rangle$, and $\langle\hat{Z}_4\rangle$ for two values of dual-species detuning ratio; $\nu=0.1$ (dashed-curves) and $\nu=0.3$ (solid-curves). $\langle\hat{Z}_1\rangle$ (panel f) falls sharply from $\approx0.55$ at $\Delta_{\text{Rb}}/\Omega_0=0$ to near zero by
$\Delta_{\text{Rb}}/\Omega_0\approx1.5$ and remains close to zero for both
values of $\nu$. $\langle\hat{Z}_2\rangle$ (panel g) develops a negative dip
$\approx-0.25$ at intermediate detuning before relaxing back toward
zero, consistent with the odd-parity behavior of $\langle\hat{Z}_2\rangle$
in panel (d). $\langle\hat{Z}_4\rangle$ (panel h), whose 11 enclosed
vertices give it odd parity, shows the same qualitative sign as
$\langle\hat{Z}_2\rangle$: for $\nu=0.3$ it develops a pronounced negative
excursion, reaching $\approx-0.35$ near $\Delta_{\text{Rb}}/\Omega_0\approx
4.5$, while for $\nu=0.1$ it instead remains close to zero. The qualitative agreement between the $N=21$ and $N=30$ columns
supports the interpretation that the parity structure of $\langle\hat{Z}\rangle$
is a genuine feature of the dimer-forming dynamics rather than a
finite-size artifact of either array.

While the diagonal operator $\hat{Z}$ reads out the classical dimer occupation along a string, detecting genuine quantum coherence between distinct dimer configurations requires an off-diagonal string operator $\hat{X}$, defined locally on each triangle that the string threads through~\cite{verresen2021prediction, semeghini2021probing, mauron2025predicting}. For a string crossing a given edge $e$ of the basis states in the reduced Hilbert space in Eq.~(\ref{Eq: Basis-States}), the associated off-diagonal local operator $\hat{X}_e$ acts as
\begin{equation}
	\hat{X}_e= \iconTriangleStringX : 
	\left\{\hskip0.5cm
	\begin{split}
		\iconTriangleA \tikz\draw[-latex,color=black, line width=1.5 pt] (0,0) to (0.75,0); - \iconTriangleB \\
		\iconTriangleB \tikz\draw[-latex,color=black, line width=1.5 pt] (0,0) to (0.75,0); - \iconTriangleA\\
		\iconTriangleC \tikz\draw[-latex,color=black, line width=1.5 pt] (0,0) to (1.2,0); \iconTriangleD\\
		\iconTriangleD \tikz\draw[-latex,color=black, line width=1.5 pt] (0,0) to (1.1,0); \iconTriangleC\\
	\end{split}
	\right.,\label{Eq: NONDiagonalOperator}
\end{equation}
\begin{figure*}[t]
	\centering
	\includegraphics[width=\textwidth]{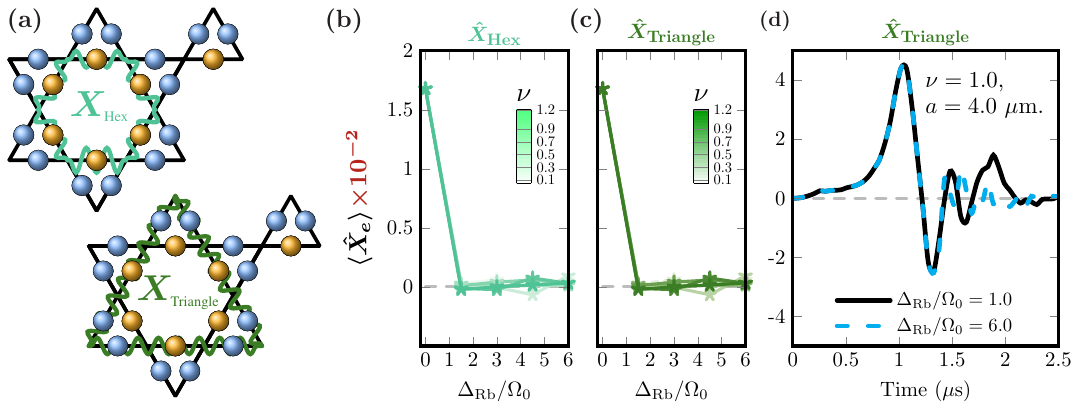}
	\caption{\textbf{Topological off-diagonal string operators}. 
		(a)The two geometrically distinct closed loops
		$\hat{X}_{\text{Hex}}$ and $\hat{X}_{\text{Triangle}}$, both enclosing
		the same 6 vertices. 
		(b,c)The evolution of $\langle\hat{X}_e\rangle$ as a
		function of $\Delta_{\text{Rb}}/\Omega_0$ for different values of the
		dual-species detuning ratio $\nu \in [0.1,1.2]$, evaluated on
		$\hat{X}_{\text{Hex}}$ and $\hat{X}_{\text{Triangle}}$, respectively.
		(d) The time-dependent evolution of $\langle\hat{X}_{\text{Triangle}}\rangle$ over
		the course of the quasi-adiabatic sweep, for $\nu=1.0$ and lattice spacing constant $a=4.0~\si{\mu m}$, comparing the trajectory toward a shallow final
		detuning ($\Delta_{\text{Rb}}/\Omega_0=1.0$, solid black) with a deep
		one ($\Delta_{\text{Rb}}/\Omega_0=6.0$, dashed cyan).
	}
	\label{Figure-TSOP-X}
\end{figure*}
i.e. $\hat{X}_e$ toggles the dimer on and off the crossed edge $e$ (with a relative $-1$ sign) while simultaneously exchanging the dimer between the two uncrossed edges $f$ and $g$ (with a relative $+1$ sign). It acts on wiggly strings by shuffling the dimers on crossed triangles~\cite{mauron2025predicting}. Each $\hat{X}_e$ is Hermitian and unitary, $\hat{X}_e^2 = \mathds{1}$, and, by construction, maps blockade-valid triangle configurations only onto other blockade-valid configurations. 

In Fig.~\ref{Figure-TSOP-X}(a), we show two closed paths for the off-diagonal operators, $\boldmath \hat{X}_{\text{Hex}}$ ($\boldmath \hat{X}_{\text{Triangle}}$), where the string crosses the edges of Hexagon (Triangle) on a Kagome lattice of $N=21$ atoms. The operator $\boldmath \hat{X}_{\text{Hex}}$ wiggles across Cs atoms only, while $\boldmath \hat{X}_{\text{Triangle}}$ wiggles twice across some triangles. Despite this difference in microscopic path, both loops enclose the
identical set of 6 vertices. In Appendix~\ref{Appendix: AlgebraString}, we prove the equivalence of these two strings algebraically. Numerically, we show the evolution of both string operators $\langle\hat{X}_{\text{Hex}}\rangle$ and $\langle\hat{X}_{\text{Triangle}}\rangle$ in Figs.~\ref{Figure-TSOP-X}(b) and (c), respectively, as a function of $\Delta_{\text{Rb}}/\Omega_{0}$ and it verifies the equivalence between the two geometrically distinct closed-loop realizations of these off-diagonal string operators. Both curves show the same qualitative behavior: $\langle\hat{X}_e\rangle$ starts at its largest value, $\approx1.7\times10^{-2}$, at
$\Delta_{\text{Rb}}/\Omega_0=0$, before falling sharply and crossing zero
near $\Delta_{\text{Rb}}/\Omega_0\approx1.5$. For $\Delta_{\text{Rb}}/\Omega_0
\geq 2.0$, the signal remains close to zero, with a small negative dip
($\approx-0.1\times10^{-2}$) near $\Delta_{\text{Rb}}/\Omega_0\approx2$--$3$
and weak oscillations of comparable magnitude persisting out to
$\Delta_{\text{Rb}}/\Omega_0=6$. The overall magnitude of
$\langle\hat{X}_e\rangle$ is more than an order of magnitude smaller than
the corresponding $\langle\hat{Z}\rangle$ signals shown in
Fig.~\ref{Figure-TSOP-Z}, indicating that whatever coherence develops
between dimer configurations connected by this hexagonal loop is weak
compared to the classical dimer-occupation signal captured by $\hat{Z}$.
Unlike the pronounced $\nu$-dependence seen for $\langle\hat{Z}_3\rangle$ in
Fig.~\ref{Figure-TSOP-Z}(e), the six curves for different $\nu$ remain
tightly bunched throughout the sweep in both panels here, suggesting that
the (weak) coherence probed by this particular loop is comparatively
insensitive to the dual-species detuning ratio.

In Fig.~\ref{Figure-TSOP-X}(d),we show the time dependence  of $\langle\hat{X}_{\text{Triangle}}\rangle$ throughout the full quasi-adiabatic sweep process at two different values of $\Delta_{Rb}/\Omega_{0}$ for $N=21$ atoms on Kagome lattice with $a=4.0$~\si{\mu m} and $\nu=1.0$ and $t_{q}=0.25$~\si{\mu s}. Both trajectories start from
$\langle\hat{X}_{\text{Triangle}}\rangle=0$, as expected for the fully
unexcited initial state $|g\rangle^{\otimes N}$, and are
\emph{indistinguishable} up to $t=\frac{1}{2}\tau+t_{q}=1.5~\si{\mu s}$ since the atoms are swept asymmetrically according to the type of the atom only after the hold time passes. Over the interval of simultaneous evolution ($t<1.5$\si{\mu s}), the
signal rises sharply to a pronounced peak,
$\langle\hat{X}_{\text{Triangle}}\rangle\approx4.5\times10^{-2}$, near $t\approx1~\si{\mu s}$, before falling through zero and reaching a
comparably sized negative minimum, $\langle\hat{X}_{\text{Triangle}}\rangle\approx-2.3\times10^{-2}$, near $t\approx1.3$~\si{\mu s}. This transient peak indicates that the loop briefly develops substantial coherence between the dimer configurations it connects during the most rapidly-varying part of the pulse, well before the system settles into its slowly-evolving final configuration.
\section{Conclusion and Outlook \label{Sec: Conclusion}}

\textbf{\textit{Conclusion}}. We considered using a dual-species array of neutral atoms on a Kagome lattice to realize the quantum spin liquid phase of matter. The array consists of Rb and Cs atoms, which are excited to Rydberg states in the regime of Rydberg blockade. We employed individual laser pulse profiles for each atomic species, along with a sweep-hold-sweep scheme, to quasi-adiabatically excite a quantum spin liquid (QSL) state in the entire system. The holding time and the values of the final positive detunings enabled us to tune the probabilities of generating the QSL state in the case of asymmetric Rydberg interactions within the dual-species array. We identified the conditions required to drive the system into a QSL state by calculating the average Rydberg density of the system, and we studied violations of the Rydberg blockade, which lead to the emergence of monomer-dimer and double-dimer configurations in the lattice. We investigated correlations between system components by calculating the correlation length and examined how it is affected by the ratio of detunings from resonance during Rydberg excitation for Rb and Cs atoms. Additionally, we analyzed correlations between regions of the system by computing the mutual information between two subsets. We observed an increase in mutual information as the size of a subset grew, for various spatial configurations of the dual-species atoms in the array. We determined the conditions under which the system follows the area law or the volume law. We explored the potential of the dual-species array for fault-tolerant quantum computation by examining the Kitaev-Preskill topological quantum entanglement entropy, and the diagonal and off-diagonal topological string operators thereby demonstrating the existence of topological order in the dual-species atomic system.

\textbf{\textit{Outlook}}. The quantum phases of matter of 2D arrays of neutral atoms, with the dual-species arrays, is a rich direction of studying many-body physics and strongly related to interesting  phenomena in condensed matter and particle physics~\cite{gonzalez2025observation} and can be utilized for implementing multiqubit quantum gates with high fidelity~\cite{doultsinos2025quantum, delakouras2025multi}. Models with holding dynamics and topological order can be used for achieving higher-fidelities of implementation of multiqubit quantum gates and performing error correction schemes. We identified the regimes of QSL state in a dual-species array which is characterized by non-uniform interaction energies which opens the way  for its experimental implementation. Our work provides, to our knowledge, the first theoretical study that systematically exploits this heteronuclear interaction asymmetry for QSL preparation and characterization on the geometrically frustrated Kagome lattice.

\begin{acknowledgments}
This work acknowledges using the computing resources of Rzhanov Institute of Semiconductor Physics SB RAS and the information and computing center of Novosibirsk State University for performing the simulations using Julia software language. 

This work was supported by the Russian Science Foundation (Grant No.~\href{https://rscf.ru/project/25-22-00730/}{25-22-00730}). 
\end{acknowledgments}

\section*{Author contributions}
A.M.F. developed sweep-hold-sweep protocol and wrote the initial draft of the manuscript. A.M.F., G.S., and J.Ch. performed numerical simulations. I.I.B. developed methodology, reviewed and edited the final draft. I.I.B., and I.I.R. guided the work and managed the resources. All authors discussed the results and contributed to writing the manuscript.

\section*{Data availability}
Datasets supporting plots in this manuscript are available upon request from the corresponding author.


\appendix
\section{Algebraic equivalence of $\hat{X}_{\text{Hex}}$ and $\hat{X}_{\text{Triangle}}$\label{Appendix: AlgebraString}}

To show that the two loops in Fig.~\ref{Figure-TSOP-X}(a) define the same
operator despite following different microscopic paths, we first establish
an algebraic identity obeyed by the single-triangle operators
$\hat{X}_e$ defined in Eq.~\eqref{Eq: NONDiagonalOperator}. Label the three edges of
a given triangle $a$, $b$, $c$, with the basis of the reduced Hilbert space
\begin{equation}
	\left\{ |\varnothing\rangle, |a\rangle, |b\rangle, |c\rangle \right\} \equiv \left\{\iconTriangleA, \iconTriangleB, \iconTriangleC, \iconTriangleD \right\}
\end{equation} 

From Eq.~\eqref{Eq: NONDiagonalOperator}, the
matrix representations of $\hat{X}_a$, $\hat{X}_b$, $\hat{X}_c$ in this
basis are
\begin{equation}
	\hat{X}_a =
	\begin{pmatrix} 0 & -1 & 0 & 0 \\ -1 & 0 & 0 & 0 \\ 0 & 0 & 0 & 1 \\ 0 & 0 & 1 & 0 \end{pmatrix},
	\label{eq:XabcMatricesA}
\end{equation}
\begin{equation}
	\hat{X}_b =
	\begin{pmatrix} 0 & 0 & -1 & 0 \\ 0 & 0 & 0 & 1 \\ -1 & 0 & 0 & 0 \\ 0 & 1 & 0 & 0 \end{pmatrix},
	\label{eq:XabcMatricesB}
\end{equation}
\begin{equation}
	\hat{X}_c =
	\begin{pmatrix} 0 & 0 & 0 & -1 \\ 0 & 0 & 1 & 0 \\ 0 & 1 & 0 & 0 \\ -1 & 0 & 0 & 0 \end{pmatrix}.
	\label{eq:XabcMatricesC}
\end{equation}
Each $\hat{X}_e$ is Hermitian and satisfies $\hat{X}_e^2 = \mathds{1}$, so
every $\hat{X}_e$ is an involution with eigenvalues $\pm1$. Direct
matrix multiplication gives
\begin{equation}
	\hat{X}_a \hat{X}_b =
	\begin{pmatrix} 0 & 0 & 0 & -1 \\ 0 & 0 & 1 & 0 \\ 0 & 1 & 0 & 0 \\ -1 & 0 & 0 & 0 \end{pmatrix}
	= \hat{X}_c
	= \hat{X}_b \hat{X}_a,
	\label{eq:XaXbXc}
\end{equation}
so that crossing two edges of a triangle ($a$ and $b$) is algebraically
identical to crossing the third, uncrossed edge ($c$) alone; the two
orderings agree, so $\hat{X}_a$ and $\hat{X}_b$ commute on this
subspace. A direct consequence is
\begin{equation}
	\hat{X}_a \hat{X}_b \hat{X}_c = \hat{X}_c^2 = \mathds{1},
	\label{eq:XaXbXcIdentity}
\end{equation}
i.e., crossing all three edges of a triangle is the identity and
contributes nothing to a string operator.

From Eqs.~\eqref{eq:XaXbXc} and \eqref{eq:XaXbXcIdentity}, let us reduce
$\hat{X}_{\text{Triangle}}$, whose path crosses two edges at several of the
six vertices it encloses as shown in Fig.~\ref{Figure-TSOP-X}(a), edge by edge. Labeling the six enclosed vertices $k=1,\dots,6$ and denoting by $e_k$ the single
edge crossed by $\hat{X}_{\text{Hex}}$ at vertex $k$, every vertex where
$\hat{X}_{\text{Triangle}}$ instead crosses the two \emph{other} edges
$\{a_k,b_k\}$ (with $\{a_k,b_k,e_k\}$ the three edges of vertex $k$) reduces
via Eq.~\eqref{eq:XaXbXc} as
\begin{equation}
	\hat{X}_{a_k}\hat{X}_{b_k} = \hat{X}_{e_k}.
	\label{eq:vertexReduction}
\end{equation}
Applying Eq.~\eqref{eq:vertexReduction} at every vertex where the two paths
differ, and noting that at the remaining vertices both loops already cross
the same single edge, gives
\begin{equation}
	\hat{X}_{\text{Triangle}} = \bigotimes_{k=1}^{6} \hat{X}_{e_k}
	= \hat{X}_{\text{Hex}},
	\label{eq:XequivProof}
\end{equation}
since both loops enclose the identical set of six vertices. The two
operators are therefore not merely topologically equivalent in a loose
sense, but algebraically identical, term by term, once each two-edge
crossing is reduced via Eq.~\eqref{eq:vertexReduction}. Consequently
$\langle\hat{X}_{\text{Triangle}}\rangle = \langle\hat{X}_{\text{Hex}}\rangle$
for any state $|\psi\rangle$, exactly as observed numerically in
Fig.~\ref{Figure-TSOP-X}(b) and \ref{Figure-TSOP-X}(c).

\nocite{*}

\bibliography{References}

\end{document}